\documentclass[letterpaper]{article}
\usepackage[preprint]{aaai2027}
\usepackage[hyphens]{url}
\usepackage{graphicx}
\usepackage{natbib}
\usepackage{caption}
\DeclareCaptionStyle{ruled}{labelfont=normalfont,labelsep=colon,strut=off}
\usepackage{microtype}
\usepackage{amsmath,amssymb}
\usepackage{booktabs}
\usepackage{subcaption}
\usepackage{cleveref}
\usepackage{multirow}
\usepackage{enumitem}
\usepackage{array}
\usepackage{pifont}
\newcommand{\cmark}{\ding{51}}
\newcommand{\xmark}{\ding{55}}

\graphicspath{{figs_final/}{figs_v2/}{figs/}}
\newcommand{\src}[1]{}

\crefname{equation}{Eq.}{Eqs.}
\Crefname{equation}{Eq.}{Eqs.}
\crefname{section}{Section}{Sections}
\Crefname{section}{Section}{Sections}
\crefname{subsection}{Section}{Sections}
\Crefname{subsection}{Section}{Sections}
\crefname{appendix}{Appendix}{Appendices}
\Crefname{appendix}{Appendix}{Appendices}
\crefname{subappendix}{Appendix}{Appendices}
\Crefname{subappendix}{Appendix}{Appendices}
\newcommand{\methodname}{\textsc{AnchorDraft}}

\title{Alignment Drift in Single-Model Speculative Decoding for ASR:\\
Mechanism, Correction, and Cost}
\author{
Xinyu Wang\textsuperscript{\rm 1,2}\thanks{Email: xinyu@boson.ai},
Huapeng Zhou\textsuperscript{\rm 1},
Ziyu Zhao\textsuperscript{\rm 2},
Silin Meng\textsuperscript{\rm 1},\\
Ke Bai\textsuperscript{\rm 1},
Dongming Shen\textsuperscript{\rm 1},
Xiao-Wen Chang\textsuperscript{\rm 2},
Alex Smola\textsuperscript{\rm 1}
}
\affiliations{
\textsuperscript{\rm 1}Boson AI\\
\textsuperscript{\rm 2}McGill University
}

\newcommand{\LadderCleanFullL}{$1.06{\to}1.28{\to}1.33$}
\newcommand{\LadderCleanHeadroom}{$6\%{\to}4\%{\to}2\%$}
\newcommand{\LadderOtherFullL}{$0.88{\to}1.11{\to}1.17$}
\newcommand{\LadderOtherHeadroom}{$17\%{\to}15\%{\to}13\%$}
\newcommand{\LadderTedFullL}{$0.91{\to}1.16{\to}1.29$}
\newcommand{\LadderTedHeadroom}{$14\%{\to}17\%{\to}14\%$}
\newcommand{\LadderGigaFullL}{$0.76{\to}1.01{\to}1.09$}
\newcommand{\LadderGigaHeadroom}{$24\%{\to}23\%{\to}23\%$}
\newcommand{\LadderFleursFullL}{$0.67{\to}1.00{\to}1.04$}
\newcommand{\LadderFleursHeadroom}{$19\%{\to}21\%{\to}17\%$}

\newcommand{\ZeroSixStdSeedZeroHardOracleMinusShift}{$0.767$}

\newcommand{\ZeroSixStdSeedOneHardOracleMinusShift}{$0.772$}

\newcommand{\CampaignSharedSliceMacro}{$0.035$}
\newcommand{\CampaignSharedSliceMacroCI}{$[+0.028, +0.041]$}
\newcommand{\CheckpointContrastMacro}{$0.732$}
\newcommand{\CheckpointContrastCI}{$[+0.706, +0.757]$}
\newcommand{\CheckpointContrastRatio}{$22\times$}
\newcommand{\FeatureNoiseZeroKOne}{$0.386\ [0.369,0.403]$}
\newcommand{\FeatureNoiseZeroKTwo}{$0.425\ [0.401,0.448]$}
\newcommand{\FeatureNoiseZeroKThree}{$0.378\ [0.340,0.416]$}
\newcommand{\FeatureNoiseZeroKFour}{$0.272\ [0.173,0.364]$}
\newcommand{\FeatureNoiseThreeKOne}{$0.432\ [0.415,0.450]$}
\newcommand{\FeatureNoiseThreeKTwo}{$0.442\ [0.419,0.464]$}
\newcommand{\FeatureNoiseThreeKThree}{$0.389\ [0.350,0.429]$}
\newcommand{\FeatureNoiseThreeKFour}{$0.286\ [0.181,0.386]$}
\newcommand{\FeatureNoiseSixKOne}{$0.434\ [0.415,0.454]$}
\newcommand{\FeatureNoiseSixKTwo}{$0.427\ [0.403,0.451]$}
\newcommand{\FeatureNoiseSixKThree}{$0.417\ [0.377,0.456]$}
\newcommand{\FeatureNoiseSixKFour}{$0.249\ [0.143,0.359]$}

\newcommand{\SupSeventeenAllSetRestartGain}{$+0.009$}
\newcommand{\SupSeventeenAllSetContinuationGain}{$+0.050$}

\newcommand{\SupTimingObserved}{$+6.8\%$}
\newcommand{\SupTimingSeedRange}{$+6.4\%$--$+7.1\%$}

\newcommand{\SupZeroSixRestartMean}{$+0.021$}
\newcommand{\SupZeroSixContinuationMean}{$+0.111$}

\newcommand{\SupTimingCI}{$[+5.8\%, +7.7\%]$}

\newcommand{\SupTimingSetRange}{$4.5$--$9.4\%$}

\newcommand{\RouteHeadroomBefore}{$0.155$}
\newcommand{\RouteHeadroomAfter}{$0.109$}
\newcommand{\RouteHeadroomDrop}{$29.7\%$}
\newcommand{\RouteHeadroomAbsDrop}{$0.046$}
\newcommand{\RouteHeadroomAbsDropCI}{$[0.030, 0.062]$}
\newcommand{\RouteHeadroomAfterCI}{$[0.087, 0.130]$}

\newcommand{\RouteFiveZeroBefore}{$0.157$}
\newcommand{\RouteFiveZeroAfter}{$0.107$}
\newcommand{\RouteFiveZeroAbsDrop}{$0.050$}
\newcommand{\RouteFiveZeroRelDrop}{$31.8\%$}
\newcommand{\RouteFiveOneBefore}{$0.153$}
\newcommand{\RouteFiveOneAfter}{$0.111$}
\newcommand{\RouteFiveOneAbsDrop}{$0.042$}
\newcommand{\RouteFiveOneRelDrop}{$27.4\%$}

\newcommand{\CampaignCleanCharged}{$-0.6\%$}
\newcommand{\CampaignCleanChargedCI}{$[-1.2\%, +0.1\%]$}

\newcommand{\CampaignCleanGross}{$+0.9\%$}
\newcommand{\CampaignCleanRestartGain}{$+0.002$}
\newcommand{\CampaignCleanContinuationGain}{$+0.015$}

\newcommand{\CampaignCleanAnchorError}{$3.5$}
\newcommand{\CampaignOtherCharged}{$+4.3\%$}
\newcommand{\CampaignOtherChargedCI}{$[+2.9\%, +5.7\%]$}

\newcommand{\CampaignOtherGross}{$+5.8\%$}
\newcommand{\CampaignOtherRestartGain}{$+0.006$}
\newcommand{\CampaignOtherContinuationGain}{$+0.075$}

\newcommand{\CampaignOtherAnchorError}{$8.2$}
\newcommand{\CampaignTedCharged}{$+3.5\%$}
\newcommand{\CampaignTedChargedCI}{$[+2.1\%, +4.9\%]$}

\newcommand{\CampaignTedGross}{$+5.0\%$}
\newcommand{\CampaignTedRestartGain}{$+0.005$}
\newcommand{\CampaignTedContinuationGain}{$+0.065$}

\newcommand{\CampaignTedAnchorError}{$9.0$}
\newcommand{\CampaignGigaCharged}{$+9.8\%$}
\newcommand{\CampaignGigaChargedCI}{$[+8.1\%, +11.5\%]$}

\newcommand{\CampaignGigaGross}{$+11.3\%$}
\newcommand{\CampaignGigaRestartGain}{$+0.010$}
\newcommand{\CampaignGigaContinuationGain}{$+0.140$}

\newcommand{\CampaignGigaAnchorError}{$18.5$}
\newcommand{\CampaignFleursCharged}{$+8.1\%$}
\newcommand{\CampaignFleursChargedCI}{$[+6.5\%, +9.7\%]$}
\newcommand{\RuntimeAllSetMean}{$+5.0\%$}
\newcommand{\RuntimeAllSetSeedRange}{$+4.7\%$--$+5.4\%$}

\newcommand{\CampaignFleursGross}{$+9.6\%$}
\newcommand{\CampaignFleursRestartGain}{$+0.009$}
\newcommand{\CampaignFleursContinuationGain}{$+0.115$}

\newcommand{\CampaignFleursAnchorError}{$15.2$}

\newcommand{\SupSeventeenAllSetTimingObserved}{$+4.5\%$}
\newcommand{\SupSeventeenAllSetTimingCI}{$[+3.55\%, +5.57\%]$}
\newcommand{\SupSeventeenAllSetTimingSeedRange}{$+3.9\%$--$+5.0\%$}

\newcommand{\SupSeventeenAllSetTimingSetRange}{$2.7$--$6.2\%$}

\newcommand{\RouteSeventeenBefore}{$0.111$}
\newcommand{\RouteSeventeenAfter}{$0.084$}
\newcommand{\RouteSeventeenDrop}{$24.3\%$}
\newcommand{\RouteSeventeenAbsDrop}{$0.027$}
\newcommand{\RouteSeventeenAbsDropCI}{$[0.018, 0.037]$}
\newcommand{\RouteSeventeenAfterCI}{$[0.063, 0.102]$}

\newcommand{\RouteSeventeenFiveZeroBefore}{$0.111$}
\newcommand{\RouteSeventeenFiveZeroAfter}{$0.084$}
\newcommand{\RouteSeventeenFiveZeroAbsDrop}{$0.027$}
\newcommand{\RouteSeventeenFiveZeroRelDrop}{$24.5\%$}
\newcommand{\RouteSeventeenFiveOneBefore}{$0.108$}
\newcommand{\RouteSeventeenFiveOneAfter}{$0.084$}
\newcommand{\RouteSeventeenFiveOneAbsDrop}{$0.024$}
\newcommand{\RouteSeventeenFiveOneRelDrop}{$22.0\%$}
\newcommand{\RouteSeventeenFiveTwoBefore}{$0.113$}
\newcommand{\RouteSeventeenFiveTwoAfter}{$0.083$}
\newcommand{\RouteSeventeenFiveTwoAbsDrop}{$0.030$}
\newcommand{\RouteSeventeenFiveTwoRelDrop}{$26.4\%$}

\newcommand{\FiveStackControlAbsoluteSpeedup}{$1.472\times$}
\newcommand{\FiveStackControlAbsoluteSpeedupCI}{$[1.41, 1.51]\times$}
\newcommand{\FiveStackAnchorAbsoluteSpeedup}{$1.531\times$}
\newcommand{\FiveStackAnchorAbsoluteSpeedupCI}{$[1.50, 1.56]\times$}
\newcommand{\FiveStackCombinedAbsoluteSpeedup}{$1.562\times$}
\newcommand{\FiveStackCombinedAbsoluteSpeedupCI}{$[1.54, 1.61]\times$}
\newcommand{\FiveStackGainOverAnchor}{$+1.8\%$}
\newcommand{\FiveStackGainOverAnchorCI}{$[+0.9\%, +2.7\%]$}
\newcommand{\FiveStackMaxWERChange}{$0.01$}

\newcommand{\RuntimeSeventeenOtherCharged}{$+2.3\%$}
\newcommand{\RuntimeSeventeenOtherCI}{$[+1.1\%, +3.5\%]$}

\newcommand{\RuntimeSeventeenTedCharged}{$+2.7\%$}
\newcommand{\RuntimeSeventeenTedCI}{$[+1.4\%, +4.0\%]$}

\newcommand{\RuntimeSeventeenGigaCharged}{$+6.8\%$}
\newcommand{\RuntimeSeventeenGigaCI}{$[+5.1\%, +8.5\%]$}

\newcommand{\RuntimeSeventeenFleursCharged}{$+5.7\%$}
\newcommand{\RuntimeSeventeenFleursCI}{$[+4.0\%, +7.4\%]$}

\newcommand{\RuntimeSeventeenCleanCharged}{$-0.4\%$}
\newcommand{\RuntimeSeventeenCleanCI}{$[-1.0\%, +0.2\%]$}

\newcommand{\RuntimeSeventeenAllSetMean}{$+3.4\%$}

\newcommand{\RuntimeSeventeenAllSetSeedRange}{$+3.0\%$--$+3.8\%$}

\newcommand{\RuntimeAcrossScaleSignMatches}{$10/10$}

\newcommand{\RuntimeSeventeenOtherAnchorError}{$6.4$}
\newcommand{\RuntimeSeventeenTedAnchorError}{$7.1$}
\newcommand{\RuntimeSeventeenGigaAnchorError}{$14.2$}
\newcommand{\RuntimeSeventeenFleursAnchorError}{$12.7$}
\newcommand{\RuntimeSeventeenOtherRestartGain}{$+0.003$}
\newcommand{\RuntimeSeventeenTedRestartGain}{$+0.003$}
\newcommand{\RuntimeSeventeenGigaRestartGain}{$+0.006$}
\newcommand{\RuntimeSeventeenFleursRestartGain}{$+0.005$}
\newcommand{\RuntimeSeventeenCleanRestartGain}{$+0.0015$}
\newcommand{\RuntimeSeventeenOtherContinuationGain}{$+0.041$}
\newcommand{\RuntimeSeventeenTedContinuationGain}{$+0.046$}
\newcommand{\RuntimeSeventeenGigaContinuationGain}{$+0.089$}
\newcommand{\RuntimeSeventeenFleursContinuationGain}{$+0.076$}
\newcommand{\RuntimeSeventeenCleanContinuationGain}{$+0.010$}

\newcommand{\RuntimeSeventeenOtherGross}{$+3.4\%$}
\newcommand{\RuntimeSeventeenTedGross}{$+3.8\%$}
\newcommand{\RuntimeSeventeenGigaGross}{$+7.9\%$}
\newcommand{\RuntimeSeventeenFleursGross}{$+6.8\%$}
\newcommand{\RuntimeSeventeenCleanGross}{$+0.7\%$}

\newcommand{\OfficialCorrectVsWrong}{$+0.254$}
\newcommand{\OfficialCorrectVsWrongCI}{$[+0.241,+0.268]$}

\newcommand{\OfficialDepthErrorSlope}{$+3.18$}
\newcommand{\OfficialDepthErrorSlopeCI}{$[+2.99,+3.37]$}
\newcommand{\OfficialDepthAcceptanceSlope}{$-0.118$}
\newcommand{\OfficialDepthAcceptanceSlopeCI}{$[-0.129,-0.106]$}
\newcommand{\OfficialDepthFourSupport}{$1.2\%$}

\newcommand{\OfficialSurvivalTwoCorrectVsFull}{$+0.105$}
\newcommand{\OfficialSurvivalTwoCorrectVsFullCI}{$[+0.099,+0.113]$}
\newcommand{\OfficialSurvivalTwoWrongVsFull}{$-0.122$}
\newcommand{\OfficialSurvivalTwoWrongVsFullCI}{$[-0.129,-0.115]$}
\newcommand{\OfficialSurvivalTwoCorrectVsWrong}{$+0.227$}
\newcommand{\OfficialSurvivalTwoCorrectVsWrongCI}{$[+0.219,+0.236]$}
\newcommand{\OfficialSurvivalThreeCorrectVsFull}{$+0.083$}
\newcommand{\OfficialSurvivalThreeCorrectVsFullCI}{$[+0.077,+0.089]$}
\newcommand{\OfficialSurvivalThreeWrongVsFull}{$-0.039$}
\newcommand{\OfficialSurvivalThreeWrongVsFullCI}{$[-0.043,-0.035]$}
\newcommand{\OfficialSurvivalThreeCorrectVsWrong}{$+0.122$}
\newcommand{\OfficialSurvivalThreeCorrectVsWrongCI}{$[+0.116,+0.128]$}
\newcommand{\OfficialLengthCorrectVsFull}{$+0.262$}
\newcommand{\OfficialLengthCorrectVsFullCI}{$[+0.248,+0.276]$}
\newcommand{\OfficialLengthWrongVsFull}{$-0.370$}
\newcommand{\OfficialLengthWrongVsFullCI}{$[-0.386,-0.356]$}
\newcommand{\OfficialLengthCorrectVsWrong}{$+0.632$}
\newcommand{\OfficialLengthCorrectVsWrongCI}{$[+0.612,+0.653]$}

\newcommand{\AltLaterContinuation}{$-0.037$}
\newcommand{\AltLaterSpeed}{$1.10\times$}
\newcommand{\AltDeepContinuation}{$\approx 0$}
\newcommand{\AltDeepSpeed}{$1.00\times$}
\newcommand{\AltKLContinuation}{$-0.111$}
\newcommand{\AltKLSpeed}{$0.99\times$}
\newcommand{\AltScheduledContinuation}{$+0.054$}
\newcommand{\AltTreeLength}{$1.36{\to}1.70$}
\newcommand{\AltTreeSpeed}{$1.60{\to}1.56\times$}

\newcommand{\SystemFullRecognizerSpeed}{$0.59$--$0.70\times$}

\newcommand{\MatchedFullCleanSpeed}{$0.592\ [0.577,0.607]$}
\newcommand{\MatchedFullOtherSpeed}{$0.654\ [0.638,0.669]$}
\newcommand{\MatchedFullTedSpeed}{$0.701\ [0.690,0.712]$}
\newcommand{\MatchedFullGigaSpeed}{$0.698\ [0.687,0.709]$}
\newcommand{\MatchedFullFleursSpeed}{$0.657\ [0.642,0.672]$}
\newcommand{\MatchedSelfCleanSpeed}{$1.526$}
\newcommand{\MatchedSelfOtherSpeed}{$1.517$}
\newcommand{\MatchedSelfTedSpeed}{$1.613$}
\newcommand{\MatchedSelfGigaSpeed}{$1.449$}
\newcommand{\MatchedSelfFleursSpeed}{$1.438$}
\newcommand{\MatchedShallowSpeedRange}{$1.44$--$1.61\times$}
\newcommand{\MatchedFullRestart}{$0.941$--$0.973$}
\newcommand{\MatchedFullContinuation}{$0.963$--$0.978$}

\newcommand{\MatchedFullParameters}{$600$M}
\newcommand{\MatchedShallowParameters}{$76.2$M}
\newcommand{\MatchedFullStaticWeight}{$1.46$ GiB}
\newcommand{\MismatchUtteranceCount}{$11/150$}
\newcommand{\MismatchTokenPositions}{$244/8{,}113$}
\newcommand{\MismatchMeanWERDelta}{$-0.193$}
\newcommand{\MismatchMeanWERDeltaCI}{$[-0.893,+0.483]$}
\newcommand{\MismatchMedianWERDelta}{$0.000$}
\newcommand{\MismatchWEROrdering}{$0.182/0.727/0.091$}
\newcommand{\PrecisionArgmaxFlipFraction}{$0.0017$}
\newcommand{\PrecisionFlipMedianMargin}{$0.00$}
\newcommand{\AnchorSourceDraftBehavior}{$2$--$20\%$}
\newcommand{\AnchorSourceRateBehavior}{$17$--$32\%$}
\newcommand{\AnchorSourceStaleBehavior}{$64$--$66\%$}
\newcommand{\AnchorSourceStaleSpeed}{$-5.3$--$-0.8\%$}
\newcommand{\AnchorSourceVerifyBehavior}{$83$--$86\%$}
\newcommand{\AnchorSourceVerifySpeed}{$+3.4\%$}

\newcommand{\BatchRhoSmall}{0.0275}

\newcommand{\OfflineAcceptedLength}{$9.4$}
\newcommand{\ChainedAcceptedLength}{$4.83$}
\newcommand{\RealLoopAcceptedLength}{$4.35$}
\newcommand{\OfflineRestartAcceptance}{$0.92$}
\newcommand{\RealLoopRestartAcceptance}{$0.635$}
\newcommand{\QwenVerificationShare}{$86$--$89\%$}
\newcommand{\QwenDraftShare}{$6.3$--$9.0\%$}

\newcommand{\AnchorTrainTotal}{$3{,}933$}

\newcommand{\AnchorZeroControlWER}{$6.420$}
\newcommand{\AnchorZeroControlAOne}{$0.731$}
\newcommand{\AnchorZeroControlContinuation}{$0.512$}
\newcommand{\AnchorZeroControlLength}{$1.105$}
\newcommand{\AnchorZeroControlLatency}{$1.000$}
\newcommand{\AnchorZeroControlSpeed}{$1.000$}
\newcommand{\AnchorZeroMethodWER}{$6.420$}
\newcommand{\AnchorZeroMethodAOne}{$0.754$}
\newcommand{\AnchorZeroMethodContinuation}{$0.628$}
\newcommand{\AnchorZeroMethodLength}{$1.228$}
\newcommand{\AnchorZeroMethodLatency}{$0.934$}
\newcommand{\AnchorZeroMethodSpeed}{$1.071$}
\newcommand{\AnchorOneControlWER}{$6.420$}
\newcommand{\AnchorOneControlAOne}{$0.728$}
\newcommand{\AnchorOneControlContinuation}{$0.518$}
\newcommand{\AnchorOneControlLength}{$1.105$}
\newcommand{\AnchorOneControlLatency}{$1.000$}
\newcommand{\AnchorOneControlSpeed}{$1.000$}
\newcommand{\AnchorOneMethodWER}{$6.430$}
\newcommand{\AnchorOneMethodAOne}{$0.747$}
\newcommand{\AnchorOneMethodContinuation}{$0.623$}
\newcommand{\AnchorOneMethodLength}{$1.212$}
\newcommand{\AnchorOneMethodLatency}{$0.940$}
\newcommand{\AnchorOneMethodSpeed}{$1.064$}
\newcommand{\AnchorSeventeenZeroControlWER}{$5.606$}
\newcommand{\AnchorSeventeenZeroControlAOne}{$0.816$}
\newcommand{\AnchorSeventeenZeroControlContinuation}{$0.679$}
\newcommand{\AnchorSeventeenZeroControlLength}{$1.371$}
\newcommand{\AnchorSeventeenZeroControlLatency}{$1.000$}
\newcommand{\AnchorSeventeenZeroControlSpeed}{$1.000$}
\newcommand{\AnchorSeventeenZeroMethodWER}{$5.608$}
\newcommand{\AnchorSeventeenZeroMethodAOne}{$0.826$}
\newcommand{\AnchorSeventeenZeroMethodContinuation}{$0.728$}
\newcommand{\AnchorSeventeenZeroMethodLength}{$1.427$}
\newcommand{\AnchorSeventeenZeroMethodLatency}{$0.956$}
\newcommand{\AnchorSeventeenZeroMethodSpeed}{$1.046$}
\newcommand{\AnchorSeventeenOneControlWER}{$5.606$}
\newcommand{\AnchorSeventeenOneControlAOne}{$0.816$}
\newcommand{\AnchorSeventeenOneControlContinuation}{$0.679$}
\newcommand{\AnchorSeventeenOneControlLength}{$1.370$}
\newcommand{\AnchorSeventeenOneControlLatency}{$1.000$}
\newcommand{\AnchorSeventeenOneControlSpeed}{$1.000$}
\newcommand{\AnchorSeventeenOneMethodWER}{$5.610$}
\newcommand{\AnchorSeventeenOneMethodAOne}{$0.824$}
\newcommand{\AnchorSeventeenOneMethodContinuation}{$0.723$}
\newcommand{\AnchorSeventeenOneMethodLength}{$1.420$}
\newcommand{\AnchorSeventeenOneMethodLatency}{$0.962$}
\newcommand{\AnchorSeventeenOneMethodSpeed}{$1.039$}
\newcommand{\AnchorSeventeenTwoControlWER}{$5.606$}
\newcommand{\AnchorSeventeenTwoControlAOne}{$0.814$}
\newcommand{\AnchorSeventeenTwoControlContinuation}{$0.676$}
\newcommand{\AnchorSeventeenTwoControlLength}{$1.364$}
\newcommand{\AnchorSeventeenTwoControlLatency}{$1.000$}
\newcommand{\AnchorSeventeenTwoControlSpeed}{$1.000$}
\newcommand{\AnchorSeventeenTwoMethodWER}{$5.608$}
\newcommand{\AnchorSeventeenTwoMethodAOne}{$0.822$}
\newcommand{\AnchorSeventeenTwoMethodContinuation}{$0.734$}
\newcommand{\AnchorSeventeenTwoMethodLength}{$1.426$}
\newcommand{\AnchorSeventeenTwoMethodLatency}{$0.953$}
\newcommand{\AnchorSeventeenTwoMethodSpeed}{$1.050$}

\makeatletter
\newcommand{\preprintlabelalias}[1]{%
  \ifcsdef{r@#1}{\csletcs{r@supp-#1}{r@#1}}{}%
  \ifcsdef{r@#1@cref}{\csletcs{r@supp-#1@cref}{r@#1@cref}}{}%
}
\makeatother

\begin{document}
\forcsvlist{\preprintlabelalias}{%
app:altdraft,app:anchordraft,app:deploychar,app:fullmatrix,%
app:hyperparam,app:landscape,app:methodb,app:mixed,app:onpolicy,%
app:paired,app:runtime-cost,app:setup,app:stdrep,app:tree,%
app:whisper-feas,app:whisper-metric,fig:method_c,fig:palign,%
tab:feature-noise,tab:headline-restart,tab:loss-sensitivity,%
tab:mismatch-wer,tab:official-depth,tab:official-outcome,%
tab:paired-position-probe,tab:proxyval,tab:qkdeploy,tab:qkdeploy-06,%
tab:stacked-correction,tab:tie-exactness}
\maketitle

\begin{abstract}
Speculative decoding speeds up generation by letting a cheap draft propose
several tokens that a target model checks in one pass. In the single-model
form, the draft is a lightweight module attached to the target rather than a
separate model. Applying this design to \textbf{A}utomatic \textbf{S}peech
\textbf{R}ecognition (ASR) introduces an extra problem. The draft can read the
whole audio at every step, yet its proposals get worse as it runs on its own.
Access is not localization. The accepted text keeps the transcript position
explicit, but the draft must also track the changing audio position. In the
primary matched comparison, per-step audio access changes the first proposal
modestly but roughly doubles later-proposal acceptance. Fixed-width
windows show that the audio position explains part of this gap. A correctly
placed window recovers
continuation, while an equally narrow window at the wrong position reduces
it. Late-draft median error reaches $21$ frames in the hardest reported
condition, while target attention during verification stays within a
$2$-frame median. We test two
ways to reduce this drift. The first reads the audio position from verification
attention and uses it to guide the next draft round. It saves time only when
the extra accepted tokens offset the readout cost. The second is
\methodname{}, which teaches the draft to track the audio position during
training without changing the inference graph. The trained draft improves
end-to-end speed at both tested target scales. These results show that ASR
self-speculation depends on token prediction, audio-position tracking, and
draft cost.
\end{abstract}

\section{Introduction}
\label{sec:intro}

\textbf{L}arge \textbf{L}anguage \textbf{M}odels (LLMs) use
\textbf{A}uto\textbf{R}egressive (AR) decoding, which requires one target-model
pass per generated token. Speculative decoding reduces the number of target
passes while returning the target's greedy output under exact arithmetic. A
cheaper draft proposes
several tokens, and the target checks them in one forward pass and accepts the
longest matching prefix
~\citep{leviathan2023fast,chen2023accelerating,xia2023specdec}. Text systems
obtain these proposals from a smaller model, auxiliary prediction heads, or
early target layers
~\citep{cai2024medusa,li2024eagle,li2025eagle3,zhang2024draftverify,elhoushi2024layerskip}.
Their speed depends on both the accepted draft length and the cost of producing
it~\citep{yan2025decoding}.

Extending this design to \textbf{A}utomatic \textbf{S}peech
\textbf{R}ecognition (ASR) adds a moving audio input to this problem. Paired
systems use a second recognizer that follows the audio independently
~\citep{gandhi2023distilwhisper,specasr2025}, but require serving another
recognizer. We instead attach a lightweight draft to the target. It reuses the
tokenizer, encoded audio, and decoder cache while the target is idle. The draft
must follow the audio and cost less than the target passes it replaces.

To locate where proposals fail, we split each speculative round. The
\emph{restart} is the first proposal after verification. Later draft-only
proposals are \emph{continuation} proposals. Per-step audio access changes
restart acceptance modestly in our primary matched comparison but roughly
doubles continuation acceptance. The main loss develops while the draft runs
without the target.

Draft cost is a separate constraint. Drafts with per-step audio access reach
$1.4$--$1.7\times$ end-to-end speedup with WER close to autoregressive
decoding. A same-family full recognizer produces almost perfect proposals but
is slower than autoregressive decoding. Useful proposals must reduce enough
target work to cover their cost.

The restart and continuation split points to information that changes between
target passes. The accepted prefix records which tokens have been generated.
Audio adds a second requirement. The draft reads the entire encoded audio at
every step, yet it must still find the frame needed for its next token. We
call this frame the \emph{audio anchor}. The text position advances one token
at a time, while the audio anchor moves by a variable number of frames because
token duration varies. Location-aware and monotonic attention address a
related moving-position problem in conventional speech decoding
~\citep{chorowski2015attention,chiu2018mocha}. Here, the position must be
maintained by a shallow draft between target passes. We use \emph{alignment
drift} for the growing distance
between the draft's attention and the audio anchor during later draft steps.

Continuation decay could also reflect weak token features, confidence errors,
or limited draft capacity. We isolate audio position by fixing the visible
window width and changing only its center. A correct-position window recovers
part of the lost continuation, while a wrong-position window reduces it. The
official-split replication gives the same ordering. Audio position explains
part of the gap, but not every rejection. Internal probes separate this
position from next-token information. In the hardest reported condition,
late-draft median error reaches $21$ frames, compared with a $2$-frame median
during target verification.

These measurements connect continuation loss to a position that verification
still estimates accurately. We test two corrections. The runtime correction
uses verification attention to position the next draft round. \methodname{}
adds position supervision during training without changing inference. Both
improve continuation on the tested Qwen checkpoints. Their end-to-end benefit
depends on whether the accepted work they recover covers their cost, so all
latency results use the real cached loop.

\begin{figure*}[t]
\centering
\includegraphics[width=\textwidth]{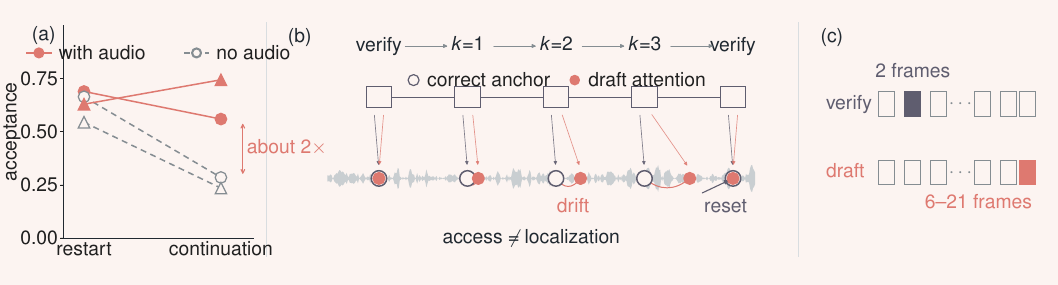}
\caption{Overview of alignment drift.
\textbf{(a)} Per-step audio access changes restart modestly and separates
continuation.
\textbf{(b)} Draft position drifts between target verifications and resets
after verification.
\textbf{(c)} Late-draft and verification anchor errors.}
\label{fig:overview}
\end{figure*}

\section{Related Work}
\label{sec:related}

\paragraph{Text Speculative Decoding.}
Lossless speculative decoding verifies several tokens from a cheaper draft
and returns the target's greedy sequence under exact arithmetic
~\citep{leviathan2023fast,chen2023accelerating}. The same draft-and-verify
pattern also applies to sequence-to-sequence generation~\citep{xia2023specdec}.
Single-model methods either attach proposal heads
~\citep{cai2024medusa,li2024eagle,li2025eagle3} or draft from early target
layers~\citep{zhang2024draftverify,elhoushi2024layerskip}. These designs avoid
serving a separate draft model, but their speed still depends on the work
accepted per round and the cost of drafting~\citep{yan2025decoding}.
MARS instead changes verification by relaxing strict token matching when the
target margin is small~\citep{song2026mars}. We keep greedy verification fixed
and study the draft state between target passes.

\paragraph{Speculative ASR.}
ASR systems draw proposals from a second recognizer~\citep{specasr2025}, an
encoder-attached lightweight decoder~\citep{lim2025hybrid}, connectionist
temporal classification outputs~\citep{saon2026selfspec}, or transcript
maps~\citep{tokenmap2025}. Distil-Whisper provides a smaller paired recognizer,
while Whisper-Medusa attaches parallel prediction heads to the recognizer
~\citep{gandhi2023distilwhisper,segalfeldman2025whispermedusa}. Partially
autoregressive masking and anticipatory recognition change the search or input
contract~\citep{okabe2025simultaneous,yusuf2024speculative}. WhisperKit
evaluates ReDrafter but removes it from one final configuration when drafting
overhead outweighs its benefit
~\citep{orhon2025whisperkit,cheng2024redrafter}. Our setting instead uses a
shallow draft attached to the target and asks why it loses proposals while
reading fixed audio memory.

\paragraph{Speech Alignment and Probes.}
Attention-based speech recognizers learn an input-output alignment
~\citep{chan2016las}. Location-aware attention, monotonic attention, and
coverage objectives constrain that alignment
~\citep{chorowski2015attention,raffel2017monotonic,chiu2018mocha,tachibana2018guided,see2017coverage}.
\methodname{} applies this idea only to the shallow draft's audio attention.
The frozen target remains unchanged, and the outcome is speculative
acceptance and latency rather than alignment quality itself. Linear probes
and logit-lens analysis can locate readable information inside the decoder,
but probe accuracy alone does not establish causal use
~\citep{tenney2019bert,hewitt2019probes,nostalgebraist2020logitlens}.
The Supplementary Material (Supp.) gives broader system comparisons and
timing decomposition in
\cref{supp-app:paired,supp-app:landscape,supp-app:deploychar}.

\section{Preliminaries and Evaluation Setup}
\label{sec:setup}

\paragraph{Speculative Loop.}
In each round, the draft proposes up to $K$ tokens, where $K$ is the maximum
draft length. The frozen target checks the block in one pass and accepts the
longest prefix that matches its greedy output. Under exact arithmetic, this
verification rule returns the same sequence as target-greedy decoding.
The first proposal after verification is the \emph{restart}. Proposals before
the next verification are \emph{continuation} proposals. Let $s_k$ be the
probability that the first $k$ proposals are accepted, with $s_0{=}1$.
Conditional acceptance at depth $k$ is $a_k^c=s_k/s_{k-1}$. We report $a_1$
as restart acceptance and $a_2^c$ as continuation acceptance. The mean number
of accepted draft tokens is $L=\sum_{k=1}^{K}s_k$. End-to-end speedup is
autoregressive greedy wall-clock divided by speculative wall-clock on the
same utterances. \textbf{W}ord \textbf{E}rror \textbf{R}ate (WER) measures
recognition error. Anchor error is the absolute distance between an attention
peak and its reference audio frame. It is reported on the post-stride audio
grid or in milliseconds. One grid frame is approximately $80$ ms for the
Qwen and Voxtral analyses in this paper.
Acceptance and accepted-length differences are absolute unless marked as
relative. We use \emph{speed gain} for the relative percentage change from a
matched control. WER is reported in percent, and WER changes are in percentage
points. Qwen aggregates weight the five evaluation sets equally within each
seed and then weight seeds equally.

\paragraph{Numerical Output Agreement.}
The verification rule is lossless under exact arithmetic, but the measured
low-precision loop is not always token-for-token identical to a separately
computed target trace. Across $150$ utterances per set, token agreement is
$0.9705$--$0.9966$, utterance agreement is $0.9267$--$0.9933$, and WER
differences are small and sign-mixed. Full-precision recomputation attributes
the differences to low-margin decisions. Supp.
\cref{supp-tab:tie-exactness,supp-tab:mismatch-wer} gives the complete check.

\paragraph{Models and Protocol.}
The draft follows the direct-token-prediction and multi-layer feature design
of the \textbf{E}xtrapolation \textbf{A}lgorithm for \textbf{G}reater
\textbf{L}anguage-model \textbf{E}fficiency
(EAGLE-3)~\citep{li2025eagle3}. We add cross-attention from each draft step to
the frozen audio encoder. We
train drafts for two Qwen3-ASR scales and Voxtral-Mini, with Whisper as a
cross-architecture comparison
~\citep{shi2026qwen3asr,liu2025voxtral,radford2023whisper}. Evaluation uses
LibriSpeech clean and other, TED-LIUM, GigaSpeech, and FLEURS
~\citep{panayotov2015librispeech,hernandez2018tedlium3,chen2021gigaspeech,conneau2022fleurs}.
The matched feasibility drafts use LibriSpeech train-clean-100. The main
causal results use official training data, development-only selection, and
official test sets. Supp.
\cref{supp-app:setup,supp-app:stdrep} give the partitions, sample sizes, and
matched configurations. Earlier checkpoints trained on prefixes from
test-partition slices are retained only as diagnostics in Supp.
\cref{supp-app:fullmatrix}.
The Qwen experiments used for the main timing results run on one NVIDIA
A100-80GB GPU at batch size one in bfloat16. Autoregressive and speculative
timing use the same utterances and start from the waveform, including
preprocessing, audio encoding, and cached decoding. The separate system
comparisons report their own hardware and timing protocol.

\paragraph{Real Cached-Loop Measurement.}
\label{sec:discipline}
Fixed-prefix scoring omits errors that enter later draft inputs. We therefore
compare offline scoring, chained survival on fixed prefixes, and the real
cached loop under one decoding setup. Mean accepted length falls from
\OfflineAcceptedLength{} under offline scoring to \ChainedAcceptedLength{}
under chained survival and \RealLoopAcceptedLength{} in the real loop.
Restart acceptance falls from \OfflineRestartAcceptance{} offline to
\RealLoopRestartAcceptance{} in the loop.
Unless explicitly labeled as offline or chained diagnostics, reported
acceptance and latency results use the real cached loop.
Supp.
\cref{supp-app:setup,supp-app:whisper-metric} give the decoding parameters
and the full comparison of measurement protocols.

\paragraph{Uncertainty.}
Unless stated otherwise, reported $95\%$ confidence intervals (CIs) use
paired utterance bootstrap resampling, with matched runs kept in each draw.
They condition on the trained checkpoints, so we report seed values or ranges
separately. Depth and state analyses use $5{,}000$ draws, while correction
timing uses $10{,}000$ and treats timing repetitions as blocks. The artifact
includes the mapping rules, bootstrap code, figure data, and run metadata.

The experiments first locate failure within the rollout, then change only the
audio position, and finally test whether the recovered work covers the cost
of correction.

\section{Alignment Drift}
\label{sec:drift}

We first locate the acceptance loss within each speculative round and show
that audio-position error grows over the same steps. We then validate the
anchor measure against an independent forced aligner and relate its error to
proposal outcomes across draft depth. Finally, we change only the position of a
fixed-width audio window to test whether position affects continuation.

\paragraph{Restart and Continuation.}
\label{sec:feas}

Per-step audio access can affect either the first proposal after verification
or the draft's ability to sustain later steps. These cases produce different
acceptance profiles. We compare matched draft variants with and without
per-step audio cross-attention. Both drafts use the same target features,
layer budget, training setup, and decoding loop. Only the draft with audio access
cross-attends to the frozen audio representation at every draft step.

\begin{table}[t]
\centering
\begingroup\small
\setlength{\tabcolsep}{3.2pt}
\begin{tabular}{@{}lcccc@{}}
\toprule
Target & Audio & $a_1\,\uparrow$ &
\shortstack{Continuation\\$a_2^c\,\uparrow$} & Speedup $\uparrow$ \\
\midrule
\multirow{2}{*}{Qwen 1.7B} & \xmark & $0.60$--$0.73$ &
$0.25$--$0.32$ & $1.14$--$1.30\times$ \\
& \cmark & $\mathbf{0.62}$--$\mathbf{0.76}$ &
$\mathbf{0.54}$--$\mathbf{0.58}$ &
$\mathbf{1.41}$--$\mathbf{1.55\times}$ \\
\midrule
\multirow{2}{*}{Voxtral 3B} & \xmark & $0.47$--$0.62$ &
$0.22$--$0.25$ & $0.95$--$1.08\times$ \\
& \cmark & $\mathbf{0.57}$--$\mathbf{0.69}$ &
$\mathbf{0.71}$--$\mathbf{0.78}$ &
$\mathbf{1.42}$--$\mathbf{1.69\times}$ \\
\bottomrule
\end{tabular}
\endgroup
\caption{Restart acceptance, continuation acceptance, and speedup for matched
drafts with and without per-step audio cross-attention. Each pair shares
the model, layer budget, LibriSpeech train-clean-100 data, and decoding loop.
Ranges give the minimum and maximum over five evaluation sets.}
\label{tab:feasibility}
\end{table}

The matched drafts differ little at restart but separate during continuation
on every evaluation set. The audio-conditioned drafts also remain faster than
autoregressive decoding. The useful effect of audio access therefore develops
between target passes rather than at the first proposal alone.
Supp. \cref{supp-tab:headline-restart} gives the five set-level
aggregates. A separate Whisper experiment follows the same ordering under an
earlier timing protocol and is reported in Supp.
\cref{supp-app:whisper-feas}.

\paragraph{Cost Boundary.}
A stronger draft tests whether complete ASR state can prevent continuation
decay. A Qwen3-ASR-0.6B recognizer drafts
for the 1.7B target with a shared tokenizer, but it runs its own encoder and
decoder at every step. It maintains nearly perfect restart and continuation
acceptance, yet reaches only \SystemFullRecognizerSpeed{} autoregressive
speed. This
control separates position maintenance from cost. The recognizer follows the
audio, but its accepted tokens do not repay its computation.

\paragraph{Official-Split Replication.}
To check whether the pattern depends on the feasibility checkpoint, we repeat
the comparison on a speaker-stratified official split. Two
independently seeded drafts with per-step audio and one matched no-audio
control use the same training and evaluation protocol. The continuation
difference remains larger than the restart difference on every evaluation
set. Its magnitude is smaller than in the feasibility runs, but its direction
is unchanged. This replication shows that the restart and continuation
ordering does not depend on the feasibility checkpoint. Supp.
\cref{supp-app:stdrep} reports the matched results.

The feasibility, official-split, and training-correction experiments use
different training sets and checkpoints. We compare directions across these
experiments but do not treat their speed values as repeated measurements of
one configuration.

\paragraph{Drift over Draft Depth.}
The restart comparison locates the loss between target passes but does not
show how it develops. We therefore follow four draft steps on two
official-split seeds and all five evaluation sets. Anchor error rises while
survival falls, with the same slope directions in every seed and set.
Cumulative survival $s_k$ tracks the fraction of initial rounds that accept
through depth $k$. Conditional acceptance $a_k^c$ shows where the curves
separate, but at later depths it compares a different survivor population in
each condition. The position intervention below reports both survival and
conditional acceptance.

\begin{figure*}[t]
\centering
\includegraphics[width=\textwidth]{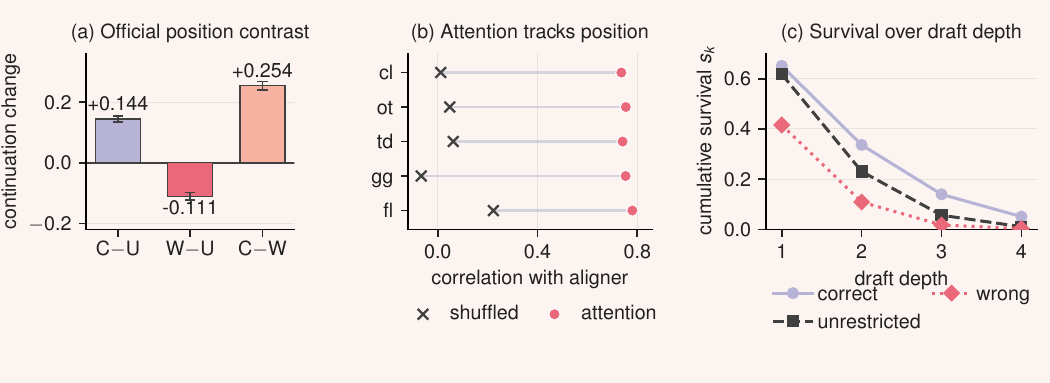}
\caption{Three measurements of audio position.
(a) Depth-two conditional-acceptance contrasts on matched official-split
examples with two seeds and $n{=}120$ utterances per evaluation set. Error bars show
paired 95\% confidence intervals. C, U, and W denote correct, unrestricted,
and shifted windows.
(b) Correlation between the draft attention peak and an independent forced
aligner, with shuffled controls.
(c) Cumulative proposal survival under the three attention conditions.}
\label{fig:mechanism}
\end{figure*}

\paragraph{Validating the Anchor Measure.}
\label{sec:anchor-validation}
Our analysis treats the peak of the draft's audio cross-attention as its
current audio position. We compare this measure with token frames from the
independent
\textbf{M}assively \textbf{M}ultilingual \textbf{S}peech \textbf{F}orced
\textbf{A}ligner (MMS-FA)~\citep{pratap2024mms}. The two positions are
correlated both across and within utterances on every evaluated set
(\cref{fig:mechanism}b).
The higher Voxtral correlations do not imply lower absolute error. Its
attention peak is systematically late relative to MMS-FA. Removing the
set-level offset substantially reduces the error. We therefore treat Voxtral
as a stable but offset boundary case, not as an example of exact
localization.
The within-utterance agreement rules out utterance duration as the only
explanation, so we use the attention peak as an operational position measure.
A paired probe at the same restart events reads more position information
after verification than from the free-running draft state. The layer and
ridge strength are fixed on separate development data. The probe shows where
position is readable but does not establish causal use. Supp.
\cref{supp-tab:proxyval,supp-tab:paired-position-probe} gives the anchor
validation and paired probe protocol.

\paragraph{Anchor Error and Proposal Outcome.}
Agreement with the independent aligner establishes that the attention peak
tracks audio position. We next compare accepted and rejected proposals at
each draft depth. Large errors are associated with rejection at restart, but
the accepted and rejected distributions overlap later in the rollout
(\cref{fig:mechanism}c). Anchor error is therefore related to proposal failure
without explaining every rejection. The intervention below tests whether
changing window position changes continuation acceptance. Supp.
\cref{supp-tab:official-outcome} gives the depth-resolved values.


\label{sec:causal}

\paragraph{Position Interventions.}
\label{sec:arms}
The prefix remains explicit during a speculative round, while the relevant
audio frame moves with every draft step. A correct-position window should
recover continuation if losing this position causes the observed decay. An
equally restrictive wrong-position window provides the matched control.
MMS-FA supplies a reference audio frame for each token, using the anchor
measure validated in \cref{sec:anchor-validation}. We compare five attention
conditions. Only the center of the draft's
cross-attention mask changes. Target verification, the accepted prefix, draft
weights, and window width remain fixed. The \emph{unrestricted} condition has
no mask. The \emph{oracle} window is centered on the current reference frame,
while the \emph{pointer} window extrapolates from previously verified frames.
The \emph{random} condition samples an unrelated center. The \emph{shift}
condition applies a fixed offset to the reference center. Random and shift are
therefore equally wide wrong-position controls. Oracle tests the effect of a
correct current position. Pointer tests whether positions from earlier
verification steps can guide the next round. Both are offline interventions,
not deployment sources. Supp.
\cref{supp-app:setup,supp-app:fullmatrix} give the window geometry, width
controls, and full intervention matrix.

These conditions yield four comparisons. Wrong-position versus unrestricted
attention measures sensitivity to position. Correct-center versus
unrestricted attention measures recoverable acceptance. Correct-center versus
an equally wide wrong-position window separates position from window width.
The errors that remain under correct positioning show how much of the failure
has another source.

\paragraph{Matched Effect of Window Position.}
The official-split replication provides the main causal check. All three arms
use the same utterances, draft weights, and target verifier. At depth two,
the correct-versus-shifted conditional-acceptance
contrast is \OfficialCorrectVsWrong{} with 95\% CI
\OfficialCorrectVsWrongCI{} (\cref{fig:mechanism}a).

The complete depth curves show where this result holds. Correct positioning
has a positive conditional-acceptance effect through depth three. At depth
four, the correct-versus-wrong interval includes zero. Only
\OfficialDepthFourSupport{} of unrestricted rounds reach that step, so the
last estimate has little support. Supp. \cref{supp-tab:official-depth}
reports survival, conditional acceptance, anchor error, and matched contrasts
at all four depths.

\begin{table}[t]
\centering
\begingroup\footnotesize
\setlength{\tabcolsep}{1.0pt}
\renewcommand{\arraystretch}{1.08}
\begin{tabular}{@{}lccc@{}}
\toprule
Outcome & C $-$ U & W $-$ U & C $-$ W \\
\midrule
$s_2$ &
\shortstack{\OfficialSurvivalTwoCorrectVsFull{}\\
\OfficialSurvivalTwoCorrectVsFullCI{}} &
\shortstack{\OfficialSurvivalTwoWrongVsFull{}\\
\OfficialSurvivalTwoWrongVsFullCI{}} &
\shortstack{\OfficialSurvivalTwoCorrectVsWrong{}\\
\OfficialSurvivalTwoCorrectVsWrongCI{}} \\
$s_3$ &
\shortstack{\OfficialSurvivalThreeCorrectVsFull{}\\
\OfficialSurvivalThreeCorrectVsFullCI{}} &
\shortstack{\OfficialSurvivalThreeWrongVsFull{}\\
\OfficialSurvivalThreeWrongVsFullCI{}} &
\shortstack{\OfficialSurvivalThreeCorrectVsWrong{}\\
\OfficialSurvivalThreeCorrectVsWrongCI{}} \\
$L$ &
\shortstack{\OfficialLengthCorrectVsFull{}\\
\OfficialLengthCorrectVsFullCI{}} &
\shortstack{\OfficialLengthWrongVsFull{}\\
\OfficialLengthWrongVsFullCI{}} &
\shortstack{\OfficialLengthCorrectVsWrong{}\\
\OfficialLengthCorrectVsWrongCI{}} \\
\bottomrule
\end{tabular}
\endgroup
\caption{Official-split position-intervention effects over two seeds and five
evaluation sets. Entries are changes in $s_k$ and
$L=\sum_{k=1}^{4}s_k$ with utterance-cluster bootstrap 95\% confidence
intervals. C, U, and W denote correct, unrestricted, and the prespecified
shifted window.}
\label{tab:survival-effects}
\end{table}

Unlike conditional acceptance, $s_k$ and $L$ start from the initial round
population. Correct centering raises both measures, while the shifted window
lowers them (\cref{tab:survival-effects}). The direction is unchanged for
either seed on every evaluation set. Audio position therefore affects
continuation. Survival remains below one under correct centering, leaving a
residual error for other sources.
Together with the independent anchor validation and depth trend, this result
supports alignment drift as one source of continuation loss.
Supp.
\cref{supp-tab:official-depth,supp-tab:official-outcome} gives the official
matched comparison and the proposal-level decomposition.
The correct-position gain persists over a broad intermediate range of window
widths. Narrow windows exclude the pointer's own error, while wide windows
re-admit distractors. Supp. \cref{supp-app:setup} reports the sweep.

\paragraph{Diagnostic Boundaries.}
\label{sec:fingerprints}
The causal result does not imply that every draft benefits from position
correction. Two measurements determine its scope. Window sensitivity is the
acceptance gap between correct and wrong centers. Drift is the growth of
anchor error under unrestricted attention.

The official-split drafts establish position sensitivity through depth three and
replicate the restart-continuation pattern. Earlier diagnostics show why
window sensitivity and drift must remain separate. One 0.6B checkpoint
responds weakly to every window, while its official-split counterpart has a
correct-versus-wrong gap across two seeds. A shared-slice replay shows that
the response depends strongly on the trained checkpoint.
We test one possible cause by retraining the official draft with three levels
of Gaussian noise added to its training features. The correct-versus-wrong
contrast remains similar at later draft depths, so feature noise does not
account for the checkpoint gap under this training recipe. Supp.
\cref{supp-app:stdrep,supp-app:fullmatrix,supp-tab:feature-noise} gives these
diagnostic comparisons.

The Voxtral diagnostic supplies a second boundary. It distinguishes correct
from wrong window centers, yet a restrictive window can reduce acceptance
even at the correct position. A useful correction therefore requires both
position sensitivity and loss of useful position information during
free-running continuation. These earlier diagnostics define the scope of the
official result rather than serving as its primary evidence.

\paragraph{Position Information inside the Decoder.}
\label{sec:layerwise}

The window intervention changes model input, but it does not show where the
decoder carries position information. We therefore train held-out linear
probes on each decoder layer. The probe predicts the
forced-alignment frame after removing the within-utterance trend with token
rank. Anchor predictability peaks in intermediate layers and weakens toward
the output across both Qwen scales and Voxtral. The paired test in
\cref{sec:anchor-validation} compares proposal-time and post-verification
states at the same restart events. These probes show where audio position is
readable, not whether the decoder uses it causally~\citep{hewitt2019probes}.
The matched window experiment tests its effect on
continuation. Supp.
\cref{supp-fig:method_c,supp-fig:palign,supp-tab:paired-position-probe} gives
the probe definitions, next-token curves, and held-out intervals.

\paragraph{Alternative Draft Changes.}
The final mechanism experiment tests simpler changes to the draft. Each
alternative changes one property of the default draft and is measured under
its own matched real-loop protocol. Later target layers change continuation
by \AltLaterContinuation{}, a two-layer draft by \AltDeepContinuation{}, and
soft KL training by \AltKLContinuation{}. Their measured speed remains
\AltKLSpeed{}--\AltLaterSpeed{}. Scheduled sampling improves continuation by
\AltScheduledContinuation{}, while tree drafting increases accepted length
but reduces speed from \AltTreeSpeed{}. None improves the quality-cost
tradeoff. Broader official-split training improves the unrestricted draft,
but a correct-position window still recovers accepted length.
Supp.
\cref{supp-app:mixed,supp-app:methodb,supp-app:onpolicy,supp-app:altdraft,supp-app:tree}
report the matched protocols and full results.

The forced-alignment pointer remains an offline intervention.
\Cref{sec:boundary} next tests a runtime correction that uses information
available during decoding and a training correction that leaves the inference
graph fixed.

\section{Correction and Cost}
\label{sec:boundary}

The position interventions identify a recoverable part of the continuation
loss. A practical correction must recover enough accepted work to offset its
cost. We study two approaches. The runtime approach obtains a new position
from each verification pass. \methodname{} instead teaches the draft to
maintain its position during training.

\begin{figure*}[t]
\centering
\includegraphics[width=\textwidth]{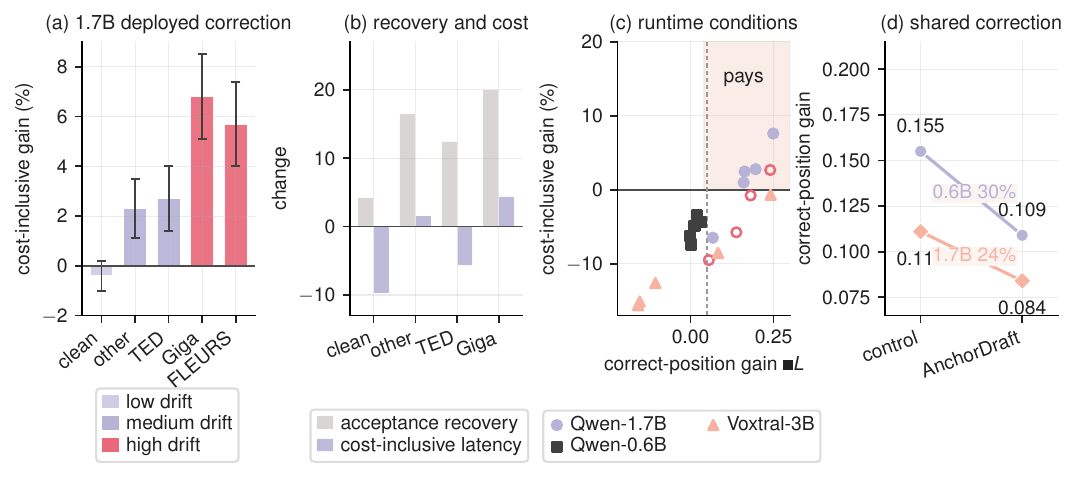}
\caption{Runtime and training-correction measurements.
(a) Runtime speed change for the $1.7$B target on five evaluation sets.
(b) Change in accepted length and cost-inclusive latency under the
correct-position intervention.
(c) Correct-position accepted-length gain and cost-inclusive pointer timing
across $19$ settings.
(d) Remaining correct-position gain before and after \methodname{}.}
\label{fig:boundary}
\end{figure*}

\paragraph{Cost Condition.}
\label{sec:runtime-cost}
Target verification accounts for \QwenVerificationShare{} of Qwen
speculative wall-clock, while drafting accounts for \QwenDraftShare{}. A
correction has room to help, but its gain must cover its own cost. Runtime
correction improves latency when
\begin{equation}
\frac{\Delta L}{1+L}\ \gtrsim\
\frac{\Delta C_{\mathrm{corr}}}{K\rho+1},
\label{eq:breakeven}
\end{equation}
where $\Delta L$ is the increase in accepted draft length,
$\Delta C_{\mathrm{corr}}$ is the correction cost relative to one target
verification, and $\rho$ is the cost of one draft step on the same scale. A
matched batch-one Qwen $1.7$B measurement gives
$\rho=\BatchRhoSmall{}$.

We use this relation as a screening rule. Position must affect prediction, the
draft must lose useful position information during continuation, and the
recovered accepted length must cover the readout cost. Across the $19$
settings in \cref{fig:boundary}c, the measured result becomes positive only
after the recovered work clears this cost. The rule also matches the direction
of \RuntimeAcrossScaleSignMatches{} deployed evaluation-set point estimates
across the two Qwen targets. This tests the sign, not the size, of the speed
gain. Voxtral fails the second condition because its position error grows
modestly while continuation remains high. Supp.
\cref{supp-app:deploychar,supp-app:fullmatrix,supp-app:runtime-cost} give the
timing decomposition, acceptance results, and evaluation-set intervals.

\paragraph{Candidate Runtime Signals.}
The intervention uses forced alignment, which is unavailable during normal
decoding. We therefore ask whether the loop already contains a usable
position estimate. We compare draft attention, speech-rate extrapolation, a
delayed position from the previous round, and verification attention. A
usable source must be accurate and current after each target pass.

A speech-rate prior reproduces only \AnchorSourceRateBehavior{} of the
correct-window behavior. The previous-round position reproduces
\AnchorSourceStaleBehavior{} but changes speed by
\AnchorSourceStaleSpeed{}. Verification attention reproduces
\AnchorSourceVerifyBehavior{} and gives a \AnchorSourceVerifySpeed{} five-set
mean speed gain. It is the only tested source that is both current and
cost-effective, although its clean interval includes zero.

\paragraph{Verification-Attention Correction.}
The target's audio attention during verification provides a fresh estimate on
the tested Qwen checkpoints. We average attention over heads at a layer
selected on development data and use its peak to center the next draft window.
The implementation reads the existing cache, uses no forced alignment, and
adds no target forward pass. The same rule selects layer $21$ for two
independently trained $1.7$B drafts.

In matched traces, verification reduces a late-draft error of up to $21$
frames to a $2$-frame median, and the next restart remains close to that
position. After charging the readout cost, the $1.7$B correction improves the
five-set mean. Its observed seed-level gains span
\RuntimeSeventeenAllSetSeedRange{}, while the clean interval includes zero.
At $0.6$B, the corresponding seed range is \RuntimeAllSetSeedRange{}.
Supp.
\cref{supp-tab:qkdeploy,supp-tab:qkdeploy-06,supp-app:stdrep} report the
layer rule, readout cost, acceptance recovery, and evaluation-set timing. The
reset and timing result show that this position is current enough to guide
the next rollout.

\paragraph{Deployment Scope.}
The runtime result above uses batch-one, short-form requests. We vary batch
size and audio duration to test how far that result extends. Target
verification amortizes faster than drafting as batch size grows, so the
draft-to-verification cost ratio rises from $0.0275$ at batch one to $0.1274$
at batch $64$. A separate real-loop sweep shows that speedup falls as audio
duration grows from roughly $18$ to $104$ seconds. The reported gains
therefore describe short, batch-one requests in the measured implementation.
Supp. \cref{supp-app:deploychar} gives both sweeps.

\paragraph{\methodname{} Training Objective.}
\label{sec:supervision}
Runtime correction reads and passes a new position after every verification.
\methodname{} instead teaches the existing draft to maintain that
position during training. It applies a Gaussian guided-attention target
~\citep{tachibana2018guided} to the shallow draft rather than the frozen
target. Each training example uses the same token sequence, target features,
and encoded audio as the control. Forced alignment maps the next token at
position $t$ to frame $c_t$. The first draft layer produces attention
$A_{t,hj}$ from head $h$ to audio index $j$. We average over heads to obtain
$\bar A_t(j)=H^{-1}\sum_h A_{t,hj}$ and match it to the normalized Gaussian
$q_t(j)\propto\exp[-(j-c_t)^2/(2\sigma^2)]$. Unaligned token positions and
padded audio keys are masked. The objective is
\begin{equation}
\mathcal L=\mathcal L_{\mathrm{tok}}+0.5\mathcal L_{\mathrm{feat}}
+\lambda\mathcal L_{\mathrm{anchor}},
\label{eq:anchordraft-main}
\end{equation}
where $\mathcal L_{\mathrm{anchor}}$ is cross-entropy between $\bar A_t$ and
$q_t$, with $\sigma>0$ and $\lambda\ge0$.

\paragraph{Training Protocol.}
Optimization updates only the draft. We select
$\lambda{=}0.1$ and $\sigma{=}5$ on a fixed holdout and train both scales on
the same \AnchorTrainTotal{}-utterance mixture of official training-set
data. The aligner and auxiliary loss are absent at inference, so \methodname{}
has the same decoding graph as its control. Target verification still accepts
only target-greedy tokens. Supp.
\cref{supp-app:anchordraft,supp-app:hyperparam} give the timestamp pairing,
masking rules, selected settings, and sensitivity sweeps.

\begin{table}[t]
\centering
\begingroup\small
\setlength{\tabcolsep}{1.5pt}
\begin{tabular}{@{}lrrr@{}}
\toprule
Target & $\Delta a_1/\Delta a_2^c\,\uparrow$ &
Speed [CI] $\uparrow$ & Set range $\uparrow$ \\
\midrule
$0.6$B &
\SupZeroSixRestartMean{}/\SupZeroSixContinuationMean{} &
\SupTimingObserved{} \SupTimingCI{} &
\SupTimingSetRange{} \\
$1.7$B &
\SupSeventeenAllSetRestartGain{}/\SupSeventeenAllSetContinuationGain{} &
\SupSeventeenAllSetTimingObserved{} \SupSeventeenAllSetTimingCI{} &
\SupSeventeenAllSetTimingSetRange{} \\
\bottomrule
\end{tabular}
\endgroup
\caption{\methodname{} changes from matched controls over five equal-weighted
evaluation sets. Set ranges contain the per-set mean speed gains.}
\label{tab:supervision}
\end{table}

\paragraph{Training Results.}
\label{sec:supervision-results}
This experiment tests whether position supervision improves the continuation
steps identified in \cref{sec:feas}. Controls match the data, optimization,
seed, and decoding graph.

\Cref{tab:supervision} shows that the gain is concentrated in continuation
rather than restart. Speed improves on every measured evaluation set, while
the maximum absolute WER change is $0.01$ percentage point.
The five-set speed gain spans \SupTimingSeedRange{} across the two $0.6$B
seeds and \SupSeventeenAllSetTimingSeedRange{} across the three $1.7$B seeds.
The post-hoc $0.6$B sensitivity sweep remains positive around the selected
weight (\cref{supp-tab:loss-sensitivity}). The Voxtral control changes by
$-0.011$ where position error remains stable during continuation.

\paragraph{Shared Mechanism.}
The overlap experiment distinguishes better position maintenance from a
general increase in draft confidence. We apply the same correct-position
intervention before and after training. The remaining gain falls at both
scales, and both reduction intervals exclude zero
(\cref{fig:boundary}d). The gain remains positive after training. Alignment
supervision therefore removes part of the position-sensitive continuation
error without replacing the runtime correction.

\paragraph{Combining the Corrections.}
The positive correct-position gain after training suggests that the runtime
readout can still recover useful work. We test this in a separate five-set
experiment at $1.7$B. The control, \methodname{}, and the stacked system reach
\FiveStackControlAbsoluteSpeedup{}, \FiveStackAnchorAbsoluteSpeedup{}, and
\FiveStackCombinedAbsoluteSpeedup{} over autoregressive decoding,
respectively. Adding runtime correction after training gives a further
\FiveStackGainOverAnchor{} gain with 95\% CI
\FiveStackGainOverAnchorCI{}, while the maximum absolute WER change remains
\FiveStackMaxWERChange{} percentage points. The corrections remain useful
together despite addressing some of the same continuation error.
Supp. \cref{supp-tab:stacked-correction} gives absolute speed
intervals.

\paragraph{System Boundary.}
We next test whether a complete recognizer provides a better quality-cost
tradeoff than the shallow draft. In the same-slice comparison, the full
recognizer adds \MatchedFullParameters{} parameters and
\MatchedFullStaticWeight{} of static weights. It accepts more tokens but
reaches only \SystemFullRecognizerSpeed{} autoregressive speed. The shallow
draft has \MatchedShallowParameters{} trainable parameters and reaches
\MatchedShallowSpeedRange{} on the same slices. Peak allocated memory remains
target-dominated in these serial batch-one runs, so the matched comparison
reports added static weight rather than a peak-memory ratio. The full
recognizer maintains position at too much cost, while the shallow draft
preserves the quality-cost balance. A separate Whisper comparison gives the
same audio-access ordering. Supp.
\cref{supp-app:anchordraft,supp-app:mixed,supp-app:paired,supp-app:whisper-feas}
give the training checks, seed-level results, and system protocols.

\FloatBarrier

\section{Conclusion and Limitations}
\label{sec:conclusion}
\label{sec:limitations}

Alignment drift is one source of continuation failure in single-model
speculative ASR. The restart and continuation split locates the loss between
target passes. Correct-position interventions recover part of the loss, while
wrong-position windows reduce acceptance. Verification attention supplies a
fresh position during decoding. \methodname{} trains the draft to maintain it
without changing inference. Both improve continuation, provided that the
recovered accepted length covers the correction cost.

The position intervention explains only part of the rejected proposals. The
study uses greedy verification that is lossless under exact arithmetic, while
low-precision computation can change a tied argmax. \methodname{} requires
token-to-frame supervision. Runtime correction is tested at two Qwen scales
under low-batch, short-form decoding with a separate readout. Other
architectures and serving conditions require new measurements. Extension to
other continuous-memory tasks remains a hypothesis.

\bibliography{custom}

\clearpage
\begin{center}
  {\LARGE\bfseries Supplementary Material\par}
  \vspace{4pt}
  {\large Alignment Drift in Single-Model Speculative Decoding for ASR:\\
  Mechanism, Correction, and Cost\par}
\end{center}
\vspace{1em}

\noindent This supplement provides the experimental protocol and supporting
evidence for the main paper. It reports the shared measurement and training
settings, internal analyses, matched position interventions, correction
results, deployment conditions, and cross-architecture comparisons.

\appendix
\section{Experimental Protocol and Exactness}
\label{app:setup}

This section specifies the evaluation scope, measurements, cached decoding,
interventions, implementation details, and numerical exactness check used
throughout the paper.

\subsection{Data and Measurement}

\paragraph{Evaluation scope.}
The experiments use drafts for two Qwen3-ASR target scales and Voxtral-Mini.
Whisper provides a cross-architecture comparison
~\citep{shi2026qwen3asr,liu2025voxtral,radford2023whisper}. Evaluation uses
LibriSpeech clean and other, TED-LIUM, GigaSpeech, and FLEURS
~\citep{panayotov2015librispeech,hernandez2018tedlium3,chen2021gigaspeech,conneau2022fleurs}.
The feasibility drafts are trained on about $6{,}000$ utterances from
LibriSpeech train-clean-100. They are evaluated on the five sets listed above.
The official-split and \methodname{} training protocols appear in
\cref{app:stdrep,app:anchordraft}. Those experiments use official test sets
and provide the causal and correction estimates. The checkpoint and
architecture analyses in
\cref{tab:fullmatrix_qwen} use separate models trained on prefixes from
test-partition slices.

\paragraph{Acceptance measures.}
Let $K$ be the maximum draft length. We write $a_1$ for acceptance of the
first proposal after verification and $a_k^c$ for acceptance at depth $k$
among rounds that reach that depth. Cumulative survival $s_k$ is the fraction
of initial rounds whose first $k$ proposals are accepted, and
$L=\sum_{k=1}^{K}s_k$ is the mean number of accepted draft tokens per round.
Conditional acceptance describes the rounds that survive to a given depth;
$s_k$ and $L$ measure the cumulative effect over all initial rounds.

\paragraph{Position, speed, and error measures.}
Anchor error is the absolute distance between an attention peak and its
reference audio frame. Frame errors use the post-stride audio grid, where one
frame is approximately $80$ ms for the Qwen and Voxtral analyses.
Cross-architecture validation in \cref{tab:proxyval,tab:misxfail} uses
milliseconds, while depth-resolved and runtime analyses use frames. Signed
offsets are labeled explicitly. Acceptance and accepted-length differences
are absolute unless marked as relative. Speed gain is the relative percentage
change from a matched control. WER is reported in percent, and WER changes are
in percentage points. For Qwen, we average the five evaluation sets equally
within each seed and then average the seeds equally.

\paragraph{Timing protocol.}
The Qwen correction measurements reported in the main paper use one NVIDIA
A100-80GB GPU at batch size one in bfloat16. Timing starts from the waveform
and includes preprocessing, audio encoding, and cached decoding.
Autoregressive and speculative runs use the same utterances, warm-up, decode
limits, and target cache. Each reported speedup is autoregressive wall-clock
divided by speculative wall-clock. Separate harnesses and hardware are
identified in \cref{app:paired}.

\paragraph{Uncertainty.}
Unless stated otherwise, brackets give $95\%$ confidence intervals (CIs).
For official-split acceptance contrasts, we resample utterances $5{,}000$
times with random seed $42$ and apply the same sampled indices to every
matched condition. These intervals measure variation across evaluation
utterances for a fixed trained checkpoint. Seed-level values or ranges
separately describe training-seed variation. The paired state probe and
\cref{tab:cimatrix} use the same $5{,}000$-draw procedure. Runtime and
\methodname{} intervals use $10{,}000$ paired utterance resamples with random
seed $20260725$. Runtime timing uses $100$ measured repetitions after $20$
warm-up runs. Control and corrected measurements are paired by utterance and
timing repetition.

\subsection{Decoding and Intervention Protocol}

\paragraph{Real cached loop.}
Offline scoring evaluates each proposal on a fixed target prefix. Chained
scoring propagates the resulting survival probabilities across depths, but it
still omits the draft states produced by earlier accepted proposals. The real
cached loop feeds accepted tokens and their cached states into the next draft
step. Unless explicitly labeled as offline or chained diagnostics, reported
acceptance and latency use the real cached loop.

\paragraph{Attention conditions.}
The unrestricted arm can attend to the full encoded audio. Correct-position,
pointer, random, and shifted arms use the same fixed-width attention mask and
differ only in its center. The correct center comes from forced alignment.
The pointer extrapolates from previously verified frames. Random and shifted
centers provide matched wrong-position controls. The pointer is an offline
intervention. The deployed correction instead reads verification attention.

\paragraph{Window geometry.}
All restricted attention conditions use a $\pm400$ ms half-width. The shifted
center moves 500 ms forward and is clipped to the valid audio grid. The random
center is sampled uniformly over that grid with a fixed seed of 0. The
$\pm400$ ms half-width was fixed before test evaluation.

\paragraph{Window-width robustness.}
The width sweep tests whether the position effect depends on the selected
half-width. On TED-LIUM and GigaSpeech, we vary the half-width while holding
the window center and decoder fixed. The correct-position gain persists over
a broad intermediate range. Narrow windows exclude pointer error, while wide
windows re-admit distractors. The position effect therefore does not depend
on the single $\pm400$ ms setting on these two evaluation sets.

\subsection{Training and Implementation}
\label{app:implementation-details}

\paragraph{Draft architecture.}
The Qwen draft combines four frozen target representations with the current
token embedding. One causal block then applies self-attention,
cross-attention to the frozen audio representation, and a feed-forward layer.
The 0.6B configuration uses hidden width 1024, 16 attention heads, and
feed-forward width 2048, for 19.2M trainable parameters. The 1.7B
configuration uses hidden width 2048, 16 heads, and feed-forward width 4096,
for 76.2M trainable parameters. Audio cross-attention is evaluated at every
draft step.

\paragraph{Alignment mapping.}
Each stored target sequence is decoded without special tokens. Its lowercase
alphanumeric text is paired with the corresponding official training
utterance by character-bigram overlap. A pair must score at least 0.45, and
extraction stops if more than 5\% of utterances remain unmatched. The
\textbf{M}assively \textbf{M}ultilingual \textbf{S}peech \textbf{F}orced
\textbf{A}ligner (MMS-FA) supplies character times~\citep{pratap2024mms}.
For each token, we decode its cumulative prefix and assign the token the time
of its last newly added alphanumeric character. Repeated tokens use their
newly decoded characters. Punctuation that adds no retained character reuses
the latest valid time. Silence is not assigned a standalone token target.
Initial special positions and tokens without a valid character alignment are
excluded from the alignment loss. Times are converted to the
utterance-specific audio grid and checked against its expected frame rate.

\paragraph{Training.}
The 3,933-utterance official training-set mixture contains 1,797 Clean, 598 Other,
598 TED-LIUM, 300 GigaSpeech, and 640 FLEURS utterances. Training runs for 45
epochs with batch size 12, AdamW at learning rate $10^{-3}$, weight decay
0.01, cosine annealing, gradient clipping at 1.0, and feature noise 0.6. The
objective combines token cross-entropy, a smooth-L1 feature loss weighted by
0.5, and the alignment loss in \cref{eq:anchordraft-supp}. A fixed holdout
selects $\lambda=0.1$ and $\sigma=5$, and the reported model uses the final
training epoch. The Gaussian target is centered on the frame for the next
token, masked at padded frames, and renormalized over valid audio positions.

\paragraph{Runtime layer.}
We evaluate layers $\{7,14,18,21,24,27\}$ on development data. The same
procedure selects layer 21 for all four evaluated Qwen heads, and this layer
remains fixed for every test set.

\subsection{Numerical Exactness}

Speculative verification reproduces target-greedy decoding under exact
arithmetic. To test whether finite-precision execution preserves this
property, we decode $150$ utterances per evaluation set with $K{=}2$, use
bfloat16 verification, and recompute the logits in float32. The bfloat16
implementation is not sequence-exact. Float32 recomputation changes the
argmax in $0.14$--$0.31\%$ of token positions, and the recorded bfloat16
margins at these positions are small.

\begin{table*}[t]
\centering
\begingroup\small
\begin{tabular}{lrrrrr}
\toprule
Set & Token exact $\uparrow$ & Utterance exact $\uparrow$ &
AR WER (\%) $\downarrow$ & Speculative WER (\%) $\downarrow$ &
Argmax flip rate $\downarrow$ \\
\midrule
Clean  & $0.9966$ & $0.9933$ & $3.52$ & $3.52$ & $0.0023$ \\
Other  & $0.9769$ & $0.9400$ & $4.03$ & $4.15$ & $0.0031$ \\
TED-LIUM & $0.9792$ & $0.9667$ & $4.63$ & $4.65$ & $0.0015$ \\
GigaSpeech & $0.9705$ & $0.9267$ & $10.95$ & $10.93$ & $0.0017$ \\
FLEURS & $0.9931$ & $0.9733$ & $4.36$ & $4.38$ & $0.0014$ \\
\bottomrule
\end{tabular}
\endgroup
\caption{Low-precision agreement with target-greedy decoding over $150$
utterances per evaluation set. The argmax flip rate compares low- and
full-precision recomputation.}
\label{tab:tie-exactness}
\end{table*}

Aggregate WER can hide degradation among the utterances whose output changes.
We therefore analyze that group on GigaSpeech, the set with the lowest
utterance agreement.

\begin{center}
\begin{minipage}{\columnwidth}
\centering
\begingroup\small
\setlength{\tabcolsep}{4pt}
\begin{tabular}{lr}
\toprule
Quantity & Value \\
\midrule
Mismatched utterances & \MismatchUtteranceCount{} \\
Mismatched token positions & \MismatchTokenPositions{} \\
Mean WER difference & \MismatchMeanWERDelta{} \\
Paired $95\%$ interval & \MismatchMeanWERDeltaCI{} \\
Median WER difference & \MismatchMedianWERDelta{} \\
Lower, equal, higher fractions & \MismatchWEROrdering{} \\
Precision argmax flip fraction & \PrecisionArgmaxFlipFraction{} \\
Flip-position median margin & \PrecisionFlipMedianMargin{} \\
\bottomrule
\end{tabular}
\endgroup
\captionof{table}{Mismatch-group WER and full-sample precision checks on
GigaSpeech. WER differences are speculative minus autoregressive in
percentage points.}
\label{tab:mismatch-wer}
\end{minipage}
\end{center}

The paired utterance bootstrap uses $5{,}000$ draws. Its interval crosses zero,
and most mismatched utterances retain the same WER. Positions where low- and
full-precision argmax differ have zero median low-precision margin. We detect
no systematic WER increase in this sample.

\paragraph{Precision propagation bound.}
Let $q$ be the probability that numerical precision changes a verified
argmax at any token, and let $n$ be the output length. If these events are
independent, then
\begin{equation}
\begin{aligned}
\Pr(\text{utterance exact}) &\ge (1-q)^n,\\
\mathbb{E}[\text{token-mismatch fraction}]
&\le \frac{n+1}{2}q .
\end{aligned}
\label{eq:tie-propagation}
\end{equation}
The first bound requires no changed argmax in the utterance. For the second,
a mismatch at position $j$ requires a changed argmax at or before $j$, whose
probability is at most $jq$. Greedy decoding can return to the same suffix, so
both expressions are conservative. This explains why a small per-token argmax
flip rate can reduce utterance-level exact match while leaving WER nearly
unchanged.

\section{Evidence for Alignment Drift}
\label{app:mechanism}

Per-step audio access matters primarily during continuation, and alignment
drift explains part of the resulting loss. This section locates the
audio-access effect within a speculative round, validates the attention-peak
anchor, examines token and position information in model states, and relates
anchor error to proposal outcome.

\subsection{Audio Access and Continuation}
\label{app:headline-restart}

To locate the effect of per-step audio access within a speculative round, we
separate the first proposal from later proposals. Matched drafts share the
target, training procedure, evaluation examples, and real cached loop. They
differ only in whether the draft uses audio cross-attention at each step.

\begin{table*}[t]
\centering
\begingroup\small
\begin{tabular}{lrrrrrr}
\toprule
Set &
\multicolumn{3}{c}{Restart acceptance $\uparrow$} &
\multicolumn{3}{c}{Continuation acceptance $\uparrow$} \\
\cmidrule(lr){2-4}\cmidrule(lr){5-7}
& With audio & Without audio & Difference &
With audio & Without audio & Difference \\
\midrule
Clean      & 0.757 & 0.730 & 0.027 & 0.540 & 0.249 & 0.291 \\
Other      & 0.739 & 0.730 & 0.009 & 0.551 & 0.258 & 0.293 \\
TED-LIUM   & 0.746 & 0.725 & 0.021 & 0.567 & 0.303 & 0.264 \\
GigaSpeech & 0.687 & 0.661 & 0.026 & 0.543 & 0.311 & 0.232 \\
FLEURS     & 0.616 & 0.597 & 0.019 & 0.575 & 0.320 & 0.255 \\
\bottomrule
\end{tabular}
\endgroup
\caption{Restart and continuation acceptance for matched drafts with and
without per-step audio access.}
\label{tab:headline-restart}
\end{table*}

Audio access changes restart acceptance modestly on every set, while the
continuation difference is much larger. The separation localizes most of the
benefit of audio access to the proposals generated between target passes.

\subsection{Validating the Attention-Peak Anchor}
\label{subsec:proxyval-method}

We measure the audio anchor as the peak of audio cross-attention. We compare
its time with an independent MMS-FA forced-alignment frame
~\citep{pratap2024mms}. Pooled correlation measures whether the two positions
move together across utterances. Within-utterance correlation removes
differences in utterance length, and absolute error measures their distance.
Shuffled targets set a reference floor. We use the peak as an operational
position measure while allowing the decoder to encode position elsewhere.

\begin{table*}[t]
\centering
\begingroup\small
\begin{tabular}{llrrr}
\toprule
Comparison & Set or feature & Pooled $r$ $\uparrow$ & Utterance $r$ $\uparrow$ &
Error in ms $\downarrow$ \\
\midrule
Qwen-1.7B peak vs.\ MMS-FA & Clean & 0.736 & 0.728 & 155 \\
Qwen-1.7B peak vs.\ MMS-FA & Other & 0.754 & 0.746 & 149 \\
Qwen-1.7B peak vs.\ MMS-FA & TED-LIUM & 0.741 & 0.750 & 142 \\
Qwen-1.7B peak vs.\ MMS-FA & FLEURS & 0.780 & 0.711 & 136 \\
Qwen-1.7B peak vs.\ MMS-FA & GigaSpeech & 0.753 & 0.756 & 144 \\
Voxtral-3B peak vs.\ MMS-FA & Clean & 0.928 & 0.932 & 230 \\
Voxtral-3B peak vs.\ MMS-FA & Other & 0.949 & 0.961 & 186 \\
\bottomrule
\end{tabular}
\endgroup
\caption{Attention-peak agreement with independent forced-alignment frames.
Errors are measured in milliseconds.}
\label{tab:proxyval}
\label{tab:perutt}
\end{table*}

The attention peak tracks the independent frame both across utterances and
within each utterance. This agreement supports its use as a measure of audio
position.

\paragraph{Voxtral offset.}
Voxtral's higher correlation does not mean that its attention peak is closer
to the forced-alignment frame. The median signed offset is $+205$ ms on Clean
and $+165$ ms on Other. Subtracting the set-level median offset reduces the
median absolute error from $230$ to $139$ ms on Clean and from $186$ to
$97$ ms on Other. All $25$ utterance-level offsets are positive on each set.
The peak therefore follows the movement of the speech position but remains
systematically later than MMS-FA.

\subsection{Token and Position Information in Model States}
\label{subsec:probe}
\label{app:curves}
\label{app:alignread}

\paragraph{Probe protocol.}
The token probe asks when the next token becomes predictable in the decoder.
At each layer, it predicts the final-layer feature and applies the frozen
output projection. GPT-2 negative log-likelihood (NLL) divides tokens into
easier and harder text-prediction groups~\citep{radford2019gpt2}. We decode
each model token with its prefix, retokenize the new text with GPT-2, and
average NLL over that text span. These groups measure text predictability
rather than linguistic or acoustic categories.

The anchor probe asks where audio position is readable. It predicts the
forced-alignment frame after removing the within-utterance trend with token
rank. Both probes use separate training and evaluation utterances, and
shuffled labels set the floor. We summarize each layerwise curve by its center
of mass (COM), which weights each layer index by its probe score.

\paragraph{Token information across layers.}
\begin{figure*}[t]
\centering
\begin{subfigure}[t]{0.48\textwidth}
  \centering
  \includegraphics[width=\linewidth]{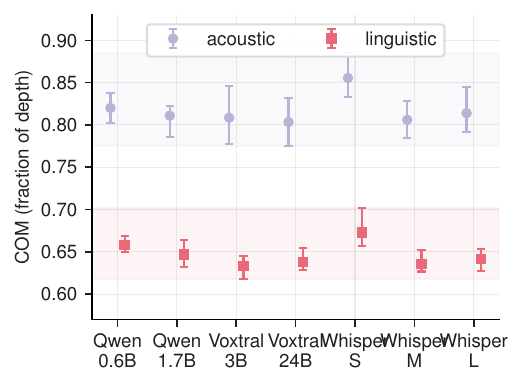}
  \caption{}
  \label{fig:combar}
\end{subfigure}\hfill
\begin{subfigure}[t]{0.48\textwidth}
  \centering
  \includegraphics[width=\linewidth]{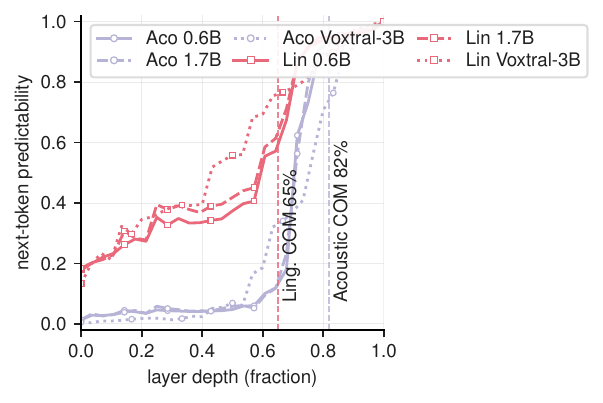}
  \caption{}
  \label{fig:method_c}
\end{subfigure}
\caption{Held-out next-token probes.
(a) Center-of-mass decoder depth for low- and high-NLL tokens.
(b) Next-token predictability by decoder layer.}
\label{fig:mech_overview}
\end{figure*}

\begin{table*}[t]
\centering
\begingroup\small
\begin{tabular}{lrr}
\toprule
Model family & High-NLL COM fraction & High-low COM gap (layers) \\
\midrule
Qwen-0.6B & 0.820 & 4.6 \\
Qwen-1.7B & 0.811 & 4.6 \\
Voxtral-3B & 0.809 & 5.3 \\
Voxtral-24B & 0.803 & 6.6 \\
Whisper-small & 0.855 & 2.2 \\
Whisper-medium & 0.806 & 4.1 \\
Whisper-large & 0.814 & 5.5 \\
\bottomrule
\end{tabular}
\endgroup
\caption{Probe summary averaged across five evaluation sets. COM fraction is
the center-of-mass layer divided by decoder depth. The gap is measured in
decoder layers.}
\label{tab:probe-summary}
\end{table*}

Across model families, tokens that are difficult to predict from text alone
have a later center-of-mass depth. The layerwise curves show the same
ordering. This experiment identifies when next-token information becomes
readable. The audio-position probe below measures a separate quantity.

\paragraph{Position information across layers.}
\label{app:palign}

The position probe measures where audio position can be decoded within the
model. At each layer, a held-out linear model predicts the forced-alignment
frame after the token-rank trend is removed.

\begin{figure}[t]
\centering
\includegraphics[width=0.92\columnwidth]{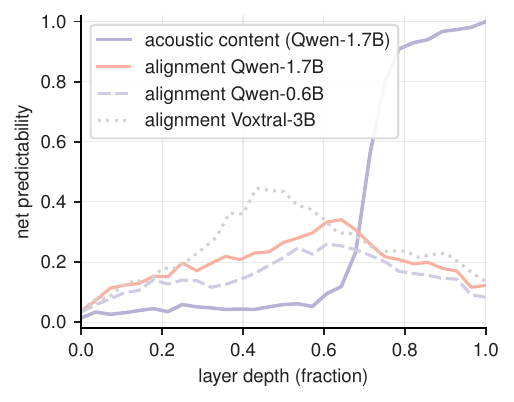}
\caption{Held-out audio-anchor probe $R^2$ by decoder layer after removing the
token-rank trend.}
\label{fig:palign}
\end{figure}

Anchor predictability peaks at layers $18$ of $28$, $17$ of $28$, and $13$
of $30$ for Qwen-1.7B, Qwen-0.6B, and Voxtral-3B. It then falls by
$64$--$70\%$ at the output. This pattern differs from the later rise of
next-token information in \cref{fig:method_c}. The two probes therefore
measure different information paths through the decoder.

\paragraph{Draft and post-verification states.}
\label{app:blindcontrast}
Verification should leave more readable position information if it restores
the draft's audio position. We compare draft and post-verification features at
the same restart event, using the same next-token alignment target, utterance
split, and bootstrap resample. Layer 20 and the ridge regularization strength
are selected on a disjoint development slice and then held fixed. Both seeds
use 600 utterances, and the held-out split contains 180 utterances.

\begin{table*}[t]
\centering
\begingroup\small
\setlength{\tabcolsep}{4.0pt}
\begin{tabular}{@{}lrrr@{}}
\toprule
Draft seed & Draft $R^2\,\uparrow$ & Post-verification $R^2\,\uparrow$ &
Paired difference $\uparrow$ \\
\midrule
bSTD &
$0.0365\ [0.0290,0.0460]$ &
$0.1956\ [0.1638,0.2328]$ &
$+0.1591\ [0.1319,0.1899]$ \\
bSTD-s1 &
$0.0359\ [0.0291,0.0446]$ &
$0.1937\ [0.1629,0.2310]$ &
$+0.1577\ [0.1310,0.1892]$ \\
Pooled & -- & -- & $+0.1584\ [0.1316,0.1892]$ \\
\bottomrule
\end{tabular}
\endgroup
\caption{Held-out position-probe results at matched restart events.
Brackets give utterance-bootstrap 95\% confidence intervals. The pooled row
reports only the paired difference.}
\label{tab:paired-position-probe}
\label{tab:blindcontrast}
\end{table*}

Post-verification features contain more linearly readable position information
than draft features in both seeds.

\subsection{Anchor Error and Proposal Outcome}
\label{subsec:misxfail}

If anchor error contributes to proposal failure, rejected proposals should
have larger errors than accepted proposals. We compare both outcomes at
matched restart positions and report the odds of rejection above a
$200$ ms error threshold.

\begin{table*}[t]
\centering
\begingroup\small
\begin{tabular}{lrrrrr}
\toprule
Set & Accepted error $\downarrow$ & Rejected error $\downarrow$ &
Large-error at accept $\downarrow$ & Large-error at reject $\downarrow$ &
Failure odds ratio $\uparrow$ \\
\midrule
Clean & 47 & 364 & 0.233 & 0.558 & 4.2 \\
Other & 48 & 208 & 0.241 & 0.507 & 3.2 \\
TED-LIUM & 46 & 211 & 0.228 & 0.509 & 3.5 \\
GigaSpeech & 45 & 236 & 0.219 & 0.517 & 3.8 \\
FLEURS & 42 & 369 & 0.175 & 0.618 & 7.6 \\
\bottomrule
\end{tabular}
\endgroup
\caption{Anchor error by proposal outcome at matched restart positions.
Errors are in milliseconds, and large error means more than $200$ ms.}
\label{tab:mislocal}
\label{tab:misxfail}
\end{table*}

Rejected proposals have larger anchor error on every evaluation set, and the
large-error rejection odds exceed one throughout the table. These
measurements link anchor error to proposal failure at restart.

\paragraph{Reconciling anchor-error summaries.}
The anchor-error summaries condition on different events.
\Cref{tab:proxyval} pools accepted and rejected restart proposals, whereas
\cref{tab:mislocal} separates restart proposals by outcome. The pooled Qwen
median of $136$--$155$ ms lies between the accepted median of $42$--$48$ ms
and the rejected median of $208$--$369$ ms. The frame-level values in the main
paper describe later draft steps in the real cached loop. The summaries differ
because they measure different parts of the rollout.

The outcome comparison is associational. The official depth-resolved analysis
in \cref{tab:official-outcome} measures the association over draft depth, and
the matched window intervention in \cref{app:fullmatrix} supplies the causal
test.

\section{Matched Position Interventions}
\label{app:fullmatrix}

Position correction applies when position sensitivity and continuation drift
co-occur. The official-split Qwen experiment below provides the primary causal
estimate. The remaining checkpoint and architecture comparisons define the
boundary of that result.

\subsection{Official-Split Position Effect}
\label{app:stdrep}

\paragraph{Official-split protocol.}
The official-split replication samples $1{,}800$ utterances from LibriSpeech
\emph{train-clean-100} by fixed-seed speaker-stratified round robin and
excludes utterances longer than $30$ seconds. Training uses $45$ epochs,
batch size $12$, learning rate $10^{-3}$, a one-layer draft, and feature noise
$0.6$. The noise is Gaussian and scaled by the empirical standard deviation
of each input feature. Two audio-conditioned seeds and one matched no-audio
draft are selected on dev-clean before evaluation on official test slices.

\paragraph{Restart and continuation replication.}
Before interpreting the position intervention, we verify the restart and
continuation ordering under matched official training and evaluation
partitions. Two audio-conditioned seeds and one no-audio control use the same
target, training recipe, and real cached loop.

\begin{table*}[t]
\centering
\begingroup\small
\begin{tabular}{lccc}
\toprule
Draft & $a_1\,\uparrow$ & Continuation $a_2^c\,\uparrow$ &
Speedup $\uparrow$ \\
\midrule
With audio, seed 0 & $0.49$--$0.72$ & $0.29$--$0.47$ &
$1.07$--$1.28\times$ \\
With audio, seed 1 & $0.50$--$0.73$ & $0.27$--$0.46$ &
$1.08$--$1.27\times$ \\
No audio & $0.43$--$0.60$ & $0.20$--$0.31$ &
$1.02$--$1.11\times$ \\
\bottomrule
\end{tabular}
\endgroup
\caption{Official-split restart, continuation, and speedup ranges over five
evaluation sets.}
\label{tab:stdrep_feas}
\end{table*}

The audio-conditioned drafts have the larger advantage during continuation
on every evaluation set. This replication places the restart and
continuation separation on the protocol used for the causal estimate.

\paragraph{Matched position intervention.}
The primary causal contrast compares correct- and wrong-position windows with
the same width. The unrestricted reference has no mask and is not
width-matched. All three conditions begin from the same rounds. We follow them
through four draft depths and compute anchor error, cumulative survival,
conditional acceptance, and rejection odds over two seeds and five evaluation
sets.

\begin{table*}[t]
\centering
\begingroup\small
\setlength{\tabcolsep}{3.0pt}
\begin{tabular}{@{}lrrrr@{}}
\toprule
Condition & Depth & Error & Survival & Conditional \\
& $k$ & mean frames $\downarrow$ & $s_k\,\uparrow$ & $a_k^c\,\uparrow$ \\
\midrule
Unrestricted & 1 & 24.98 & 0.618 & 0.618 \\
& 2 & 30.70 & 0.231 & 0.371 \\
& 3 & 33.47 & 0.057 & 0.270 \\
& 4 & 34.66 & 0.011 & 0.258 \\
\addlinespace
Correct window & 1 & 2.65 & 0.651 & 0.651 \\
& 2 & 2.90 & 0.336 & 0.520 \\
& 3 & 3.13 & 0.140 & 0.442 \\
& 4 & 3.19 & 0.051 & 0.409 \\
\addlinespace
Wrong window & 1 & 3.89 & 0.416 & 0.416 \\
& 2 & 4.01 & 0.109 & 0.265 \\
& 3 & 4.22 & 0.018 & 0.226 \\
& 4 & 4.24 & 0.004 & 0.319 \\
\bottomrule
\end{tabular}
\par\vspace{5pt}
\begin{tabular}{@{}rrrr@{}}
\toprule
Depth & Correct $-$ unrestricted & Wrong $-$ unrestricted &
Correct $-$ wrong \\
\midrule
1 & $+0.030\ [0.024,0.036]$ & $-0.206\ [-0.215,-0.197]$ &
$+0.236\ [0.227,0.245]$ \\
2 & $+0.144\ [0.133,0.155]$ & $-0.111\ [-0.122,-0.098]$ &
$+0.254\ [0.241,0.268]$ \\
3 & $+0.166\ [0.146,0.186]$ & $-0.050\ [-0.078,-0.022]$ &
$+0.216\ [0.189,0.243]$ \\
4 & $+0.135\ [0.089,0.180]$ & $+0.051\ [-0.029,0.161]$ &
$+0.084\ [-0.018,0.160]$ \\
\bottomrule
\end{tabular}

\par\vspace{5pt}
\begin{tabular}{@{}rrrr@{}}
\toprule
Depth & Accepted error & Rejected error &
\shortstack{Rejection\\odds ratio $\downarrow$} \\
& median frames $\downarrow$ & median frames $\downarrow$ & above 200 ms \\
\midrule
1 & 9.22 & 20.94 & $3.67\ [3.08,4.35]$ \\
2 & 18.11 & 22.00 & $2.02\ [1.77,2.30]$ \\
3 & 24.33 & 24.67 & $1.73\ [1.34,2.22]$ \\
4 & 27.06 & 26.22 & $1.43\ [0.73,2.67]$ \\
\bottomrule
\end{tabular}
\endgroup
\caption{Official-split depth summaries over two seeds and five sets with
$n{=}120$ utterances per set. The matched contrasts use paired
utterance-cluster resampling, and brackets give 95\% confidence intervals.}
\label{tab:stdrep_arms}
\label{tab:official-depth}
\label{tab:official-outcome}
\end{table*}

\paragraph{Depth-resolved effects.}
The unrestricted error slope is \OfficialDepthErrorSlope{} frames per step
with interval \OfficialDepthErrorSlopeCI{}. Its conditional-acceptance slope
is \OfficialDepthAcceptanceSlope{} per step with interval
\OfficialDepthAcceptanceSlopeCI{}. Both trends have the same direction in
every seed and evaluation set. The matched window contrasts exclude zero
through depth three.

\paragraph{Depth-four boundary.}
Only \OfficialDepthFourSupport{} of unrestricted rounds and $0.4\%$ of
wrong-window rounds reach depth four. At this depth, the correct-versus-wrong
interval includes zero, and accepted and rejected errors converge. The
position effect is clearest in the well-populated part of the rollout.

\subsection{Checkpoint and Architecture Comparisons}

\paragraph{Five-arm checkpoint matrix.}
These comparisons test whether the measured response to position is shared
across checkpoints and model families. The five attention conditions keep the
visible width fixed and change the window center or remove the window
restriction. The resulting trajectories separate position sensitivity from
the loss of position during continuation.

\begin{figure*}[t]
\centering
\includegraphics[width=\textwidth]{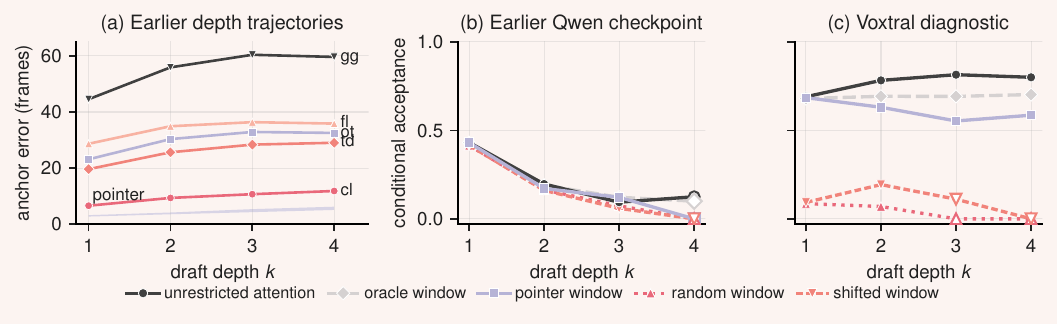}
\caption{Depth-resolved checkpoint comparisons.
(a) Anchor error for unrestricted attention and the pointer range over five
evaluation sets.
(b--c) Conditional acceptance under five attention conditions for Qwen and
Voxtral.
Open markers indicate depths reached by fewer than $2\%$ of rounds.}
\label{fig:diagnostic-depth}
\end{figure*}

\begin{table*}[t]
\centering
\begingroup\small
\setlength{\tabcolsep}{3.0pt}
\begin{tabular}{lccccc}
\toprule
Set & Unrestricted & Correct & Pointer & Random & Shift \\
\midrule
\multicolumn{6}{l}{Qwen-1.7B single-set draft} \\
Clean & 0.726 / 0.421 & 0.732 / 0.501 & 0.742 / 0.454 & 0.487 / 0.335 & 0.418 / 0.292 \\
Other & 0.598 / 0.331 & 0.655 / 0.515 & 0.654 / 0.502 & 0.394 / 0.236 & 0.392 / 0.227 \\
TED-LIUM & 0.587 / 0.344 & 0.645 / 0.478 & 0.642 / 0.447 & 0.398 / 0.209 & 0.386 / 0.212 \\
GigaSpeech & 0.520 / 0.237 & 0.615 / 0.450 & 0.607 / 0.425 & 0.348 / 0.135 & 0.368 / 0.185 \\
FLEURS & 0.483 / 0.313 & 0.535 / 0.482 & 0.526 / 0.464 & 0.329 / 0.231 & 0.316 / 0.253 \\
\midrule
\multicolumn{6}{l}{Qwen-1.7B mixed-set draft} \\
Clean & 0.817 / 0.628 & 0.830 / 0.670 & 0.823 / 0.620 & 0.500 / 0.384 & 0.361 / 0.402 \\
Other & 0.729 / 0.549 & 0.765 / 0.714 & 0.771 / 0.668 & 0.327 / 0.248 & 0.355 / 0.290 \\
TED-LIUM & 0.851 / 0.603 & 0.870 / 0.728 & 0.862 / 0.659 & 0.430 / 0.249 & 0.405 / 0.279 \\
GigaSpeech & 0.757 / 0.499 & 0.809 / 0.700 & 0.803 / 0.650 & 0.352 / 0.176 & 0.371 / 0.259 \\
\midrule
\multicolumn{6}{l}{Qwen-0.6B checkpoint comparison} \\
Clean & 0.527 / 0.319 & 0.530 / 0.315 & 0.525 / 0.312 & 0.502 / 0.259 & 0.501 / 0.301 \\
Other & 0.450 / 0.244 & 0.466 / 0.270 & 0.462 / 0.258 & 0.424 / 0.219 & 0.433 / 0.240 \\
TED-LIUM & 0.432 / 0.197 & 0.432 / 0.190 & 0.431 / 0.172 & 0.418 / 0.163 & 0.416 / 0.159 \\
GigaSpeech & 0.384 / 0.182 & 0.394 / 0.201 & 0.393 / 0.191 & 0.371 / 0.173 & 0.379 / 0.177 \\
FLEURS & 0.374 / 0.225 & 0.377 / 0.244 & 0.376 / 0.242 & 0.352 / 0.210 & 0.357 / 0.213 \\
\bottomrule
\end{tabular}
\endgroup
\caption{Five-arm acceptance matrix for the Qwen checkpoint comparisons at
$K{=}2$. Each entry gives $a_1/s_2$, and the mixed-set block reports its
four held-out sets.}
\label{tab:fullmatrix_qwen}
\end{table*}

\paragraph{Checkpoint sensitivity on a shared slice.}
The 0.6B comparison checkpoint is less sensitive to position than the
official-split checkpoint. To separate checkpoint choice from target scale,
we compare the benefit of a correctly centered window over a shifted window
on the same difficult utterances. The comparison checkpoint gives
\CampaignSharedSliceMacro{} with interval \CampaignSharedSliceMacroCI{}. The
two official-split seeds give \ZeroSixStdSeedZeroHardOracleMinusShift{} and
\ZeroSixStdSeedOneHardOracleMinusShift{}, about \CheckpointContrastRatio{}
larger. The paired official-minus-comparison difference is
\CheckpointContrastMacro{} with interval \CheckpointContrastCI{}. Target
scale, evaluation utterances, and intervention are fixed, so the difference
comes from the trained checkpoint.

\paragraph{Feature-noise control.}
Feature noise is one candidate explanation for the checkpoint gap. We retrain
the official draft at three noise levels while keeping the other training
conditions fixed.

\begin{center}
\begin{minipage}{\columnwidth}
\centering
\begingroup\small
\setlength{\tabcolsep}{5pt}
\begin{tabular}{ccc}
\toprule
Training noise & Depth & Correct minus wrong $[95\%\,\mathrm{CI}]$ \\
\midrule
\multirow{4}{*}{0.0}
& 1 & \FeatureNoiseZeroKOne{} \\
& 2 & \FeatureNoiseZeroKTwo{} \\
& 3 & \FeatureNoiseZeroKThree{} \\
& 4 & \FeatureNoiseZeroKFour{} \\
\addlinespace
\multirow{4}{*}{0.3}
& 1 & \FeatureNoiseThreeKOne{} \\
& 2 & \FeatureNoiseThreeKTwo{} \\
& 3 & \FeatureNoiseThreeKThree{} \\
& 4 & \FeatureNoiseThreeKFour{} \\
\addlinespace
\multirow{4}{*}{0.6}
& 1 & \FeatureNoiseSixKOne{} \\
& 2 & \FeatureNoiseSixKTwo{} \\
& 3 & \FeatureNoiseSixKThree{} \\
& 4 & \FeatureNoiseSixKFour{} \\
\bottomrule
\end{tabular}
\endgroup
\captionof{table}{Correct-minus-wrong conditional acceptance under matched
feature-noise training. Brackets give paired $95\%$ bootstrap intervals.
Each noise level uses one training seed.}
\label{tab:feature-noise}
\end{minipage}
\end{center}

The three drafts share the training data, architecture, optimizer, learning
rate, epoch count, and training seed. Only the Gaussian feature-noise
multiplier changes. At depths two through four, the intervals overlap
substantially. At depth one, removing noise lowers the contrast by about
$0.05$, far less than the cross-checkpoint difference. Feature noise therefore
accounts for little of the checkpoint gap under this training recipe.

\paragraph{Cross-architecture response.}
The final comparison asks whether sensitivity to window position and loss of
position during continuation occur together across model families. Within
each model, the five attention conditions use matched utterances and window
widths.

\begin{table*}[t]
\centering
\begingroup\small
\newcommand{\cino}[2]{\shortstack{$#1$\\$[#2]$}}
\newcommand{\cistar}[2]{\shortstack{$#1^*$\\$[#2]$}}
\renewcommand{\arraystretch}{1.18}
\begin{tabular}{llcccc}
\toprule
Model & Set & Correct $-$ unrestricted &
Pointer $-$ unrestricted & Pointer $-$ correct &
Random $-$ unrestricted \\
\midrule
Qwen-1.7B & Clean & \cino{+0.006}{-0.017,+0.029} & \cino{+0.012}{-0.009,+0.034} & \cistar{-0.047}{-0.081,-0.016} & \cistar{-0.261}{-0.301,-0.222} \\
Qwen-1.7B & Other & \cistar{+0.057}{+0.039,+0.075} & \cistar{+0.056}{+0.037,+0.075} & \cino{-0.009}{-0.024,+0.006} & \cistar{-0.217}{-0.240,-0.194} \\
Qwen-1.7B & TED-LIUM & \cistar{+0.063}{+0.045,+0.082} & \cistar{+0.054}{+0.035,+0.073} & \cistar{-0.028}{-0.046,-0.010} & \cistar{-0.193}{-0.217,-0.168} \\
Qwen-1.7B & GigaSpeech & \cistar{+0.099}{+0.083,+0.115} & \cistar{+0.092}{+0.078,+0.107} & \cistar{-0.021}{-0.033,-0.010} & \cistar{-0.177}{-0.193,-0.162} \\
Qwen-1.7B & FLEURS & \cistar{+0.060}{+0.039,+0.082} & \cistar{+0.048}{+0.028,+0.069} & \cino{-0.017}{-0.034,+0.000} & \cistar{-0.169}{-0.191,-0.147} \\
\midrule
Voxtral-3B & Clean & \cistar{+0.091}{+0.059,+0.124} & \cistar{+0.098}{+0.065,+0.133} & \cistar{-0.061}{-0.081,-0.043} & \cistar{-0.606}{-0.643,-0.571} \\
Voxtral-3B & Other & \cino{+0.024}{-0.007,+0.056} & \cino{-0.002}{-0.034,+0.031} & \cistar{-0.087}{-0.109,-0.063} & \cistar{-0.653}{-0.692,-0.615} \\
Voxtral-3B & TED-LIUM & \cistar{-0.042}{-0.066,-0.018} & \cistar{-0.036}{-0.061,-0.012} & \cistar{-0.052}{-0.078,-0.027} & \cistar{-0.677}{-0.708,-0.644} \\
Voxtral-3B & GigaSpeech & \cistar{-0.043}{-0.066,-0.020} & \cistar{-0.060}{-0.081,-0.039} & \cistar{-0.073}{-0.095,-0.052} & \cistar{-0.633}{-0.660,-0.607} \\
Voxtral-3B & FLEURS & \cistar{-0.054}{-0.083,-0.025} & \cistar{-0.053}{-0.081,-0.027} & \cistar{-0.028}{-0.048,-0.008} & \cistar{-0.589}{-0.626,-0.553} \\
\bottomrule
\end{tabular}
\endgroup
\caption{Paired $95\%$ utterance-cluster bootstrap intervals for the
checkpoint comparisons. Contrasts use restart acceptance except pointer minus
correct, which uses depth-two survival. A star marks an interval that excludes
zero. All intervals are pointwise.}
\label{tab:cimatrix}
\label{tab:voxreconcile}
\end{table*}

Correct placement recovers acceptance for the drifting Qwen draft. Voxtral
distinguishes correct from wrong centers, but restricting attention can reduce
acceptance even at the correct center. Its attention peak remains stable
across draft depth despite its offset from MMS-FA. Position sensitivity and
loss of position during continuation are therefore separate properties.

\section{Correction and Cost}
\label{app:deployment}

We test two corrections to the continuation loss. The runtime correction reads
a fresh position after verification and helps only when the recovered accepted
work covers its readout cost. \methodname{} teaches the existing draft to
maintain position during training without changing the inference graph. This
section reports the matched acceptance and timing results for both
corrections.

\subsection{Cost Condition}

\paragraph{Cost derivation.}
Normalize one target verification to unit cost and let one draft step cost
$\rho$. A round with maximum draft length $K$ costs approximately
$1+K\rho$ and advances $1+L$ output tokens. After correction, its cost and
progress become $1+K\rho+\Delta C_{\mathrm{corr}}$ and
$1+L+\Delta L$. The corrected round is faster when
\begin{equation}
\frac{1+K\rho+\Delta C_{\mathrm{corr}}}{1+L+\Delta L}
<
\frac{1+K\rho}{1+L}.
\end{equation}
Cross-multiplication gives
\begin{equation}
\frac{\Delta L}{1+L}
>
\frac{\Delta C_{\mathrm{corr}}}{1+K\rho}.
\end{equation}
The main paper uses this approximation to predict the sign of the speed
change.

\subsection{Runtime Correction}
\label{app:anchorbound}

\paragraph{Runtime anchor construction.}
The runtime source averages target audio attention over heads at a decoder
layer selected on development data. We choose the earliest layer whose
attention falls inside the reference window within $0.02$ of the best
candidate. Both independently trained 1.7B drafts select layer $21$ from
$\{7,14,18,21,24,27\}$. Inference recomputes the selected attention scores
from the existing key cache. It uses no forced alignment and no additional
target forward pass.
\label{app:anchor-select}

\paragraph{Required anchor accuracy.}
The runtime source must be accurate enough for its accepted-length gain to
cover its cost. We estimate this requirement using the pointer and
shifted-window arms as the correct and wrong endpoints. A parameter-free
mixture predicts accepted length for noisy anchors. Across the tested noise
levels, predicted and observed accepted length differ by at most $0.023$.

\paragraph{Candidate position sources.}
We then compare four inexpensive position sources. On the matched Other and
GigaSpeech diagnostics, draft attention reproduces
\AnchorSourceDraftBehavior{} of correct-window behavior, and the speech-rate
prior reproduces \AnchorSourceRateBehavior{}. A previous-round frame reaches
\AnchorSourceStaleBehavior{}, but its cost-inclusive result remains
\AnchorSourceStaleSpeed{}. Verification attention reaches
\AnchorSourceVerifyBehavior{} and is the only tested source that clears the
cost-inclusive threshold.

\paragraph{Measured position error.}
Matched cached-loop traces compare draft attention with verification
attention. Late-draft median error is $6$, $10$, and $21$ frames on
Clean, TED-LIUM, and GigaSpeech.
Verification attention has a $2$-frame median and a $6$-frame 90th
percentile. The result identifies verification attention as the more accurate
measured source.

\paragraph{End-to-end runtime correction.}
We apply the same correction at both Qwen scales to measure whether the
position estimate improves end-to-end speed after including its readout cost.
Each run uses a matched uncorrected loop, and we report acceptance changes
together with measured net speed.

\begin{table*}[t]
\centering
\begingroup\small
\setlength{\tabcolsep}{1.4pt}
\begin{tabular}{llrrrrr}
\toprule
Scale & Set & Error $\downarrow$ & $\Delta a_1\,\uparrow$ &
$\Delta a_2^c\,\uparrow$ & Gross $\uparrow$ &
Net [95\% CI] $\uparrow$ \\
\midrule
\multirow{5}{*}{1.7B}
& Clean & $2.75$ &
\RuntimeSeventeenCleanRestartGain{} & \RuntimeSeventeenCleanContinuationGain{} &
\RuntimeSeventeenCleanGross{} &
\RuntimeSeventeenCleanCharged{} \RuntimeSeventeenCleanCI{} \\
& Other & \RuntimeSeventeenOtherAnchorError{} &
\RuntimeSeventeenOtherRestartGain{} & \RuntimeSeventeenOtherContinuationGain{} &
\RuntimeSeventeenOtherGross{} &
\RuntimeSeventeenOtherCharged{} \RuntimeSeventeenOtherCI{} \\
& TED-LIUM & \RuntimeSeventeenTedAnchorError{} &
\RuntimeSeventeenTedRestartGain{} & \RuntimeSeventeenTedContinuationGain{} &
\RuntimeSeventeenTedGross{} &
\RuntimeSeventeenTedCharged{} \RuntimeSeventeenTedCI{} \\
& GigaSpeech & \RuntimeSeventeenGigaAnchorError{} &
\RuntimeSeventeenGigaRestartGain{} & \RuntimeSeventeenGigaContinuationGain{} &
\RuntimeSeventeenGigaGross{} &
\RuntimeSeventeenGigaCharged{} \RuntimeSeventeenGigaCI{} \\
& FLEURS & \RuntimeSeventeenFleursAnchorError{} &
\RuntimeSeventeenFleursRestartGain{} & \RuntimeSeventeenFleursContinuationGain{} &
\RuntimeSeventeenFleursGross{} &
\RuntimeSeventeenFleursCharged{} \RuntimeSeventeenFleursCI{} \\
\midrule
\multirow{5}{*}{0.6B}
& Clean & \CampaignCleanAnchorError{} & \CampaignCleanRestartGain{} &
\CampaignCleanContinuationGain{} & \CampaignCleanGross{} &
\CampaignCleanCharged{} \CampaignCleanChargedCI{} \\
& Other & \CampaignOtherAnchorError{} & \CampaignOtherRestartGain{} &
\CampaignOtherContinuationGain{} & \CampaignOtherGross{} &
\CampaignOtherCharged{} \CampaignOtherChargedCI{} \\
& TED-LIUM & \CampaignTedAnchorError{} & \CampaignTedRestartGain{} &
\CampaignTedContinuationGain{} & \CampaignTedGross{} &
\CampaignTedCharged{} \CampaignTedChargedCI{} \\
& GigaSpeech & \CampaignGigaAnchorError{} & \CampaignGigaRestartGain{} &
\CampaignGigaContinuationGain{} & \CampaignGigaGross{} &
\CampaignGigaCharged{} \CampaignGigaChargedCI{} \\
& FLEURS & \CampaignFleursAnchorError{} & \CampaignFleursRestartGain{} &
\CampaignFleursContinuationGain{} & \CampaignFleursGross{} &
\CampaignFleursCharged{} \CampaignFleursChargedCI{} \\
\bottomrule
\end{tabular}
\endgroup
\caption{Runtime correction by target scale and evaluation set. Anchor error
is in frames. Net gain is measured end to end. Gross saving is an accounting
estimate. Restart and continuation columns are absolute acceptance changes.}
\label{tab:tgtattn}
\label{tab:qkdeploy}
\label{tab:qkdeploy-mechanism}
\label{tab:qkdeploy-06}
\label{tab:qkdeploy-06-mechanism}
\end{table*}

The five-set net gains are \RuntimeSeventeenAllSetMean{} at 1.7B and
\RuntimeAllSetMean{} at 0.6B. At each scale, four of five set-level intervals
are positive. The measured cost rule matches all
\RuntimeAcrossScaleSignMatches{} evaluation-set and scale point-estimate
signs. Correction pays where the accepted work recovered in the table exceeds
the readout cost.

\subsection{Training Correction with \methodname{}}
\label{app:anchordraft}

\paragraph{Alignment objective.}
\methodname{} adds a training-only loss that places the draft's first-layer
audio attention near the forced-alignment frame for the next token. The token
and feature losses remain unchanged.

Each training example uses the same token sequence, target features, and
encoded audio as its control. At draft position $t$, the draft predicts token
$t{+}1$ and its target feature, while forced alignment supplies frame $c_t$
for that next token. With valid audio keys $\mathcal K_u$, valid token positions
$\mathcal V$, and first-layer audio attention $A_{t,hj}$, the objective is
\begin{equation}
\begin{aligned}
\bar A_t(j)&=H^{-1}\sum_h A_{t,hj},\\
q_t(j)&=\frac{\exp[-(j-c_t)^2/(2\sigma^2)]}
 {\sum_{\ell\in\mathcal K_u}\exp[-(\ell-c_t)^2/(2\sigma^2)]},\\
\mathcal L&=\mathcal L_{\mathrm{tok}}+0.5\mathcal L_{\mathrm{feat}}
-\frac{\lambda}{|\mathcal V|}\sum_{t\in\mathcal V}
 \sum_{j\in\mathcal K_u}q_t(j)\log\bar A_t(j).
\end{aligned}
\label{eq:anchordraft-supp}
\end{equation}

\paragraph{Matched training and evaluation.}
The selected setting uses $\lambda{=}0.1$ and $\sigma{=}5$ on a fixed
holdout. Both scales train on the same \AnchorTrainTotal{}-utterance
official training-set mixture. Only the draft is updated. The inference graph
is unchanged. The aligner and alignment loss are absent at inference, and
target verification keeps the same greedy output.
The matched controls use this official training-set mixture. The main
audio-access comparison uses LibriSpeech-only drafts.

To isolate the alignment objective, control and \methodname{} use the same
data, optimizer, training steps, seed, and decoding protocol at each scale.
We measure continuation, target output, and end-to-end speed with the
inference graph held fixed.

\begin{table*}[t]
\centering
\begingroup\small
\setlength{\tabcolsep}{2.2pt}
\begin{tabular}{cclrrrrrr}
\toprule
Target & Seed & Draft & WER (\%) $\downarrow$ & $a_1\,\uparrow$ & $a_2^c\,\uparrow$ &
\shortstack{Two-step\\length $\uparrow$} & Latency $\downarrow$ & Speed $\uparrow$ \\
\midrule
$0.6$B & 0 & Control & \AnchorZeroControlWER{} & \AnchorZeroControlAOne{} & \AnchorZeroControlContinuation{} & \AnchorZeroControlLength{} & \AnchorZeroControlLatency{} & \AnchorZeroControlSpeed{} \\
$0.6$B & 0 & \methodname{} & \AnchorZeroMethodWER{} & \AnchorZeroMethodAOne{} & \AnchorZeroMethodContinuation{} & \AnchorZeroMethodLength{} & \AnchorZeroMethodLatency{} & \AnchorZeroMethodSpeed{} \\
$0.6$B & 1 & Control & \AnchorOneControlWER{} & \AnchorOneControlAOne{} & \AnchorOneControlContinuation{} & \AnchorOneControlLength{} & \AnchorOneControlLatency{} & \AnchorOneControlSpeed{} \\
$0.6$B & 1 & \methodname{} & \AnchorOneMethodWER{} & \AnchorOneMethodAOne{} & \AnchorOneMethodContinuation{} & \AnchorOneMethodLength{} & \AnchorOneMethodLatency{} & \AnchorOneMethodSpeed{} \\
\midrule
$1.7$B & 0 & Control & \AnchorSeventeenZeroControlWER{} & \AnchorSeventeenZeroControlAOne{} & \AnchorSeventeenZeroControlContinuation{} & \AnchorSeventeenZeroControlLength{} & \AnchorSeventeenZeroControlLatency{} & \AnchorSeventeenZeroControlSpeed{} \\
$1.7$B & 0 & \methodname{} & \AnchorSeventeenZeroMethodWER{} & \AnchorSeventeenZeroMethodAOne{} & \AnchorSeventeenZeroMethodContinuation{} & \AnchorSeventeenZeroMethodLength{} & \AnchorSeventeenZeroMethodLatency{} & \AnchorSeventeenZeroMethodSpeed{} \\
$1.7$B & 1 & Control & \AnchorSeventeenOneControlWER{} & \AnchorSeventeenOneControlAOne{} & \AnchorSeventeenOneControlContinuation{} & \AnchorSeventeenOneControlLength{} & \AnchorSeventeenOneControlLatency{} & \AnchorSeventeenOneControlSpeed{} \\
$1.7$B & 1 & \methodname{} & \AnchorSeventeenOneMethodWER{} & \AnchorSeventeenOneMethodAOne{} & \AnchorSeventeenOneMethodContinuation{} & \AnchorSeventeenOneMethodLength{} & \AnchorSeventeenOneMethodLatency{} & \AnchorSeventeenOneMethodSpeed{} \\
$1.7$B & 2 & Control & \AnchorSeventeenTwoControlWER{} & \AnchorSeventeenTwoControlAOne{} & \AnchorSeventeenTwoControlContinuation{} & \AnchorSeventeenTwoControlLength{} & \AnchorSeventeenTwoControlLatency{} & \AnchorSeventeenTwoControlSpeed{} \\
$1.7$B & 2 & \methodname{} & \AnchorSeventeenTwoMethodWER{} & \AnchorSeventeenTwoMethodAOne{} & \AnchorSeventeenTwoMethodContinuation{} & \AnchorSeventeenTwoMethodLength{} & \AnchorSeventeenTwoMethodLatency{} & \AnchorSeventeenTwoMethodSpeed{} \\
\bottomrule
\end{tabular}
\endgroup
\caption{\methodname{} and matched controls averaged over five evaluation
sets. Two-step length is $a_1(1+a_2^c)$. Latency and speed are normalized to
the control for each seed.}
\label{tab:anchordraft_full}
\end{table*}

\paragraph{Correction results.}
Continuation increases more than restart at both scales, while WER changes by
at most $0.01$ percentage points. Five-set speed gains span
\SupTimingSeedRange{} across the two $0.6$B seeds and
\SupSeventeenAllSetTimingSeedRange{} across the three $1.7$B seeds. The result
places the training effect in the part of the rollout identified by the
restart and continuation diagnostic.

\paragraph{Cross-architecture boundary.}
The preregistered Voxtral control uses a head whose position error remains
stable during continuation. Position supervision changes its hard-domain
equal-weighted continuation-acceptance average by $-0.011$.

\paragraph{Post-hoc loss-weight sensitivity.}
We vary the alignment-loss weight to measure local sensitivity around the
selected setting. The sweep was run after the fixed holdout selected
$\lambda{=}0.1$ and did not affect the selected loss weight, checkpoint, or
test protocol.

\begin{table}[t]
\centering
\begingroup\small
\begin{tabular}{lr}
\toprule
$\lambda$ & 0.6B five-set speed gain $\uparrow$ \\
\midrule
0.00 & 0.0\% \\
0.05 & 5.6\% \\
0.10 & 6.8\% \\
0.20 & 7.0\% \\
0.30 & 6.1\% \\
\bottomrule
\end{tabular}
\endgroup
\caption{Post-hoc five-set speed sensitivity to the alignment-loss weight at
$0.6$B.}
\label{tab:loss-sensitivity}
\label{tab:speed-sensitivity}
\end{table}

Speed remains positive near the selected weight, peaks at $\lambda{=}0.2$,
and falls at $\lambda{=}0.3$.

\subsection{Relationship Between the Corrections}

\paragraph{Correction overlap.}
If \methodname{} reduces the same position-sensitive error, a
correct-position window should recover less accepted length after training.
We measure this residual gain with the window width, evaluation sets, and
aggregation fixed.

\begin{table*}[t]
\centering
\begingroup\small
\begin{tabular}{@{}clrrrr@{}}
\toprule
Target & Seed & Before $\downarrow$ & After $\downarrow$ &
Absolute reduction $\uparrow$ & Relative reduction $\uparrow$ \\
\midrule
$0.6$B & 0 & \RouteFiveZeroBefore{} & \RouteFiveZeroAfter{} & \RouteFiveZeroAbsDrop{} & \RouteFiveZeroRelDrop{} \\
$0.6$B & 1 & \RouteFiveOneBefore{} & \RouteFiveOneAfter{} & \RouteFiveOneAbsDrop{} & \RouteFiveOneRelDrop{} \\
$0.6$B & Mean & \RouteHeadroomBefore{} & \RouteHeadroomAfter{} & \RouteHeadroomAbsDrop{} & \RouteHeadroomDrop{} \\
\midrule
$1.7$B & 0 & \RouteSeventeenFiveZeroBefore{} & \RouteSeventeenFiveZeroAfter{} & \RouteSeventeenFiveZeroAbsDrop{} & \RouteSeventeenFiveZeroRelDrop{} \\
$1.7$B & 1 & \RouteSeventeenFiveOneBefore{} & \RouteSeventeenFiveOneAfter{} & \RouteSeventeenFiveOneAbsDrop{} & \RouteSeventeenFiveOneRelDrop{} \\
$1.7$B & 2 & \RouteSeventeenFiveTwoBefore{} & \RouteSeventeenFiveTwoAfter{} & \RouteSeventeenFiveTwoAbsDrop{} & \RouteSeventeenFiveTwoRelDrop{} \\
$1.7$B & Mean & \RouteSeventeenBefore{} & \RouteSeventeenAfter{} & \RouteSeventeenAbsDrop{} & \RouteSeventeenDrop{} \\
\bottomrule
\end{tabular}
\endgroup
\caption{Correct-position accepted-length gain before and after \methodname{}
over five evaluation sets.}
\label{tab:correction-overlap}
\end{table*}

The absolute-reduction intervals are \RouteHeadroomAbsDropCI{} at 0.6B and
\RouteSeventeenAbsDropCI{} at 1.7B. The remaining-gain intervals are
\RouteHeadroomAfterCI{} and \RouteSeventeenAfterCI{}. The correct-position
gain decreases after training but remains positive at both scales.
\methodname{} therefore removes part, but not all, of the error recovered by
the correct-position intervention. This establishes residual
position-sensitive headroom; the stacked experiment below tests whether the
runtime readout can use it.

\paragraph{Stacked correction.}
A final matched comparison applies runtime correction after \methodname{}
training. At 1.7B, we compare its absolute speedup with the control and
\methodname{} alone to measure the remaining runtime gain.

\begin{table}[t]
\centering
\begingroup\small
\setlength{\tabcolsep}{1.0pt}
\begin{tabular}{@{}lcc@{}}
\toprule
Configuration & Speedup vs.\ AR $\uparrow$ & 95\% CI \\
\midrule
Control & \FiveStackControlAbsoluteSpeedup{} &
\FiveStackControlAbsoluteSpeedupCI{} \\
\methodname{} & \FiveStackAnchorAbsoluteSpeedup{} &
\FiveStackAnchorAbsoluteSpeedupCI{} \\
Stacked & \FiveStackCombinedAbsoluteSpeedup{} &
\FiveStackCombinedAbsoluteSpeedupCI{} \\
\bottomrule
\end{tabular}
\endgroup
\caption{Five-set absolute speedups in the separate stacked-correction
experiment at 1.7B. The five evaluation sets receive equal weight.}
\label{tab:stacked-correction}
\end{table}

Runtime correction adds \FiveStackGainOverAnchor{} over \methodname{} with
95\% CI \FiveStackGainOverAnchorCI{}. The maximum absolute WER change is
\FiveStackMaxWERChange{} percentage points. The positive added gain shows
that runtime correction still recovers work after training.

\section{Configuration, Alternatives, and Deployment Scope}
\label{app:runtime-cost}

Continuation quality alone does not determine speed. This section tests that
boundary through draft configuration, training coverage, alternative draft
changes, and serving conditions that change the quality-cost balance.

\subsection{Draft Length and Confidence Threshold}
\label{app:hyperparam}

Let $\rho$ be the cost of one draft step relative to one target pass. A longer
draft helps only when the added survival probability covers its added
drafting cost. The batch-one Qwen $1.7$B measurement gives
$\rho=\BatchRhoSmall{}$. The real cached-loop sweep below measures that
tradeoff directly.

\begin{table}[t]
\centering
\begingroup\small
\setlength{\tabcolsep}{3.5pt}
\begin{tabular}{ccrrrr}
\toprule
$K$ & Threshold & Speedup $\uparrow$ & $a_1\,\uparrow$ &
$a_2^c\,\uparrow$ & $s_2\,\uparrow$ \\
\midrule
1 & 0.0 & 1.217 & 0.770 & -- & -- \\
\textbf{2} & \textbf{0.0} & \textbf{1.332} & 0.750 & 0.477 & 0.358 \\
3 & 0.0 & 1.320 & 0.755 & 0.482 & 0.364 \\
4 & 0.0 & 1.280 & 0.753 & 0.482 & 0.363 \\
5 & 0.0 & 1.247 & 0.753 & 0.478 & 0.360 \\
8 & 0.0 & 1.129 & 0.753 & 0.478 & 0.360 \\
\midrule
1 & 0.3 & 1.201 & 0.770 & -- & -- \\
2 & 0.3 & 1.352 & 0.753 & 0.465 & 0.350 \\
\textbf{3} & \textbf{0.3} & \textbf{1.365} & 0.753 & 0.465 & 0.350 \\
4 & 0.3 & 1.332 & 0.751 & 0.465 & 0.349 \\
5 & 0.3 & 1.347 & 0.751 & 0.462 & 0.347 \\
8 & 0.3 & 1.326 & 0.751 & 0.462 & 0.347 \\
\bottomrule
\end{tabular}
\endgroup
\caption{Draft-length and confidence-threshold sweep for Qwen-1.7B on $80$
clean utterances. A dash marks an undefined continuation metric.}
\label{tab:gamma_conf}
\end{table}

\begin{figure}[t]
\centering
\includegraphics[width=\columnwidth]{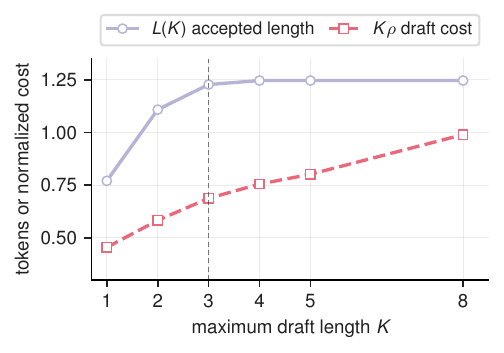}
\caption{Accepted length and per-round draft cost by maximum draft length
$K$.}
\label{fig:lgamma_overhead}
\end{figure}

With threshold $0.0$, speed peaks at $K{=}2$ ($1.332\times$). Threshold $0.3$
shifts the measured peak to $K{=}3$ ($1.365\times$). In both sweeps, accepted
length saturates while draft cost continues to grow. These optima apply to the
$80$ Clean utterances and batch-one implementation measured here.

\subsection{Training Coverage}
\label{app:mixed}

The training-coverage experiment asks whether broader data alone removes the
correct-position gap. We compare clean-only, three-set, and five-set drafts
under the same official-split protocol, then measure unrestricted accepted
length and the additional gain from a correct-position window.

\begin{table*}[t]
\centering
\begingroup\small
\setlength{\tabcolsep}{6.0pt}
\begin{tabular}{lcrr}
\toprule
Set & In three-set mix &
\shortstack{Unrestricted $L\,\uparrow$\\Clean-only $\to$ 3-set $\to$ 5-set} &
\shortstack{Relative correct-position gap $\downarrow$\\
Clean-only $\to$ 3-set $\to$ 5-set} \\
\midrule
Clean       & \cmark & \LadderCleanFullL{}  & \LadderCleanHeadroom{} \\
Other       & \cmark & \LadderOtherFullL{}  & \LadderOtherHeadroom{} \\
FLEURS      & \cmark & \LadderFleursFullL{} & \LadderFleursHeadroom{} \\
TED-LIUM    & \xmark & \LadderTedFullL{}     & \LadderTedHeadroom{} \\
GigaSpeech  & \xmark & \LadderGigaFullL{}    & \LadderGigaHeadroom{} \\
\bottomrule
\end{tabular}
\endgroup
\caption{Training-coverage comparison. The relative correct-position gap is
$(L_{\mathrm{correct}}-L_{\mathrm{unrestricted}})/
L_{\mathrm{unrestricted}}$. All runs use $K{=}2$ and $n{=}120$ per
evaluation set.}
\label{tab:stdmix}
\end{table*}

Broader training raises unrestricted accepted length, but a correct-position
window still helps when the evaluation set is represented in training.
Training coverage alone therefore does not remove the position gap.

\subsection{Controls for Alternative Explanations}
\label{app:methodb}
\label{app:onpolicy}
\label{app:lossy}
\label{app:altdraft}
\label{app:tree}

Training coverage is evaluated in the preceding subsection. The controls here
test whether continuation loss can be resolved by token information, draft
capacity, the training objective, candidate coverage, or a relaxed output
constraint. Each control changes one property of the default draft under its
matched evaluation protocol. We compare its acceptance and speed with the
corresponding default.

\begin{table*}[t]
\centering
\begingroup\small
\begin{tabular}{p{2.6cm}p{2.8cm}p{3.3cm}p{2.5cm}p{3.6cm}}
\toprule
Alternative & Intended target & Acceptance effect & Speed effect &
Measured outcome \\
\midrule
Later target layers & Token information & \AltLaterContinuation{} continuation &
\AltLaterSpeed{} & Continuation falls \\
Two-layer draft & Draft capacity & Approximately no continuation change &
\AltDeepSpeed{} & Added cost removes the gain \\
Soft KL objective & Token quality & \AltKLContinuation{} continuation &
\AltKLSpeed{} & Worse than cross-entropy \\
Scheduled sampling & Continuation exposure &
\AltScheduledContinuation{} conditional continuation & Not timed &
Restart and two-step survival fall \\
Tree draft & Candidate coverage & $L$ changes by \AltTreeLength{} &
\AltTreeSpeed{} & Wider verification removes the gain \\
Relaxed tokens & Acceptance & Small WER-neutral increase & Narrow gain &
Changes the output constraint \\
\bottomrule
\end{tabular}
\endgroup
\caption{Alternative draft and verification changes under their matched
evaluation protocols.}
\label{tab:alternative-corrections}
\label{tab:methodb_full}
\label{tab:onpolicy}
\label{tab:tree}
\end{table*}

\paragraph{Tree-decoder implementation check.}
This control tests whether proposing several token paths can improve
continuation through wider candidate coverage. The draft builds a fixed
candidate tree, and the target verifies all nodes in one pass with a tree
attention mask. A one-branch tree reproduces the linear decoder token for
token. On the matched long-audio test, the $11$-node tree raises root hit rate
from $0.76$ to $0.88$.

Under the tested budgets, the default one-layer cross-entropy draft gives the
best measured tradeoff among the argmax-preserving alternatives. Added
capacity, later feature access, and wider candidate coverage do not by
themselves resolve continuation loss in these comparisons.

\subsection{Deployment Conditions}
\label{app:deploychar}

These measurements test whether the batch-one, short-form timing result
extends to other serving conditions. CUDA events separate drafting and target
verification, a microbenchmark varies batch size at fixed accepted length,
and a real-loop sweep concatenates audio into longer inputs.

\begin{figure*}[t]
\centering
\begin{subfigure}[t]{0.32\textwidth}
\centering
\includegraphics[width=\linewidth]{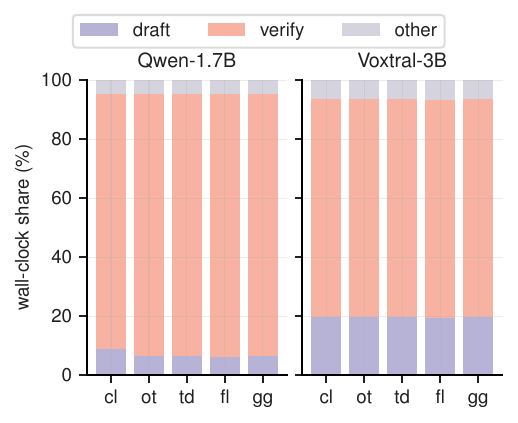}
\caption{Wall-clock components}
\label{fig:decomp}
\end{subfigure}\hfill
\begin{subfigure}[t]{0.32\textwidth}
\centering
\includegraphics[width=\linewidth]{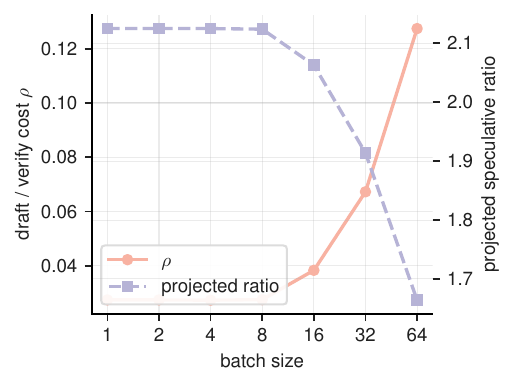}
\caption{Batch size}
\label{fig:batchrho}
\end{subfigure}\hfill
\begin{subfigure}[t]{0.32\textwidth}
\centering
\includegraphics[width=\linewidth]{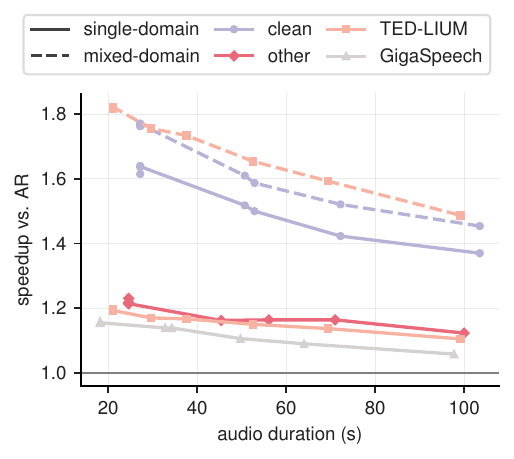}
\caption{Audio duration}
\label{fig:longform}
\end{subfigure}
\caption{Deployment measurements.
(a) Wall-clock composition.
(b) Draft-to-verification cost ratio and projected speed by batch size.
(c) Real-loop speedup by concatenated-audio duration.}
\label{fig:deployment-scope}
\end{figure*}

\paragraph{Wall-clock composition.}
CUDA-event measurements assign \QwenVerificationShare{} of Qwen wall-clock to
target verification and \QwenDraftShare{} to drafting. The corresponding
Voxtral draft share is $19.4$--$19.7\%$.

\paragraph{Batch-size scaling.}
The batch sweep shows that verification amortizes faster than the shallow
draft, increasing $\rho$ at larger batch sizes.

\paragraph{Audio-duration scaling.}
Across six draft and evaluation-set configurations, speedup falls as
concatenated audio grows from roughly $18$ to $104$ seconds.

\paragraph{Measured deployment scope.}
These measurements characterize the short-form, batch-one operating regime
used for the reported timing results.

\section{Cross-Architecture and System Comparisons}
\label{app:paired}

Proposal quality must repay draft cost. This section tests that condition with
a same-family recognizer, checks the audio-access ordering in Whisper, and
places the repository measurements beside published systems.

\subsection{Same-Family Recognizer as Draft}

\paragraph{Matched setup.}
A complete recognizer can maintain its own audio state, but its computation
may erase the benefit of accepted tokens. We test this tradeoff with a
Qwen3-ASR-0.6B recognizer that drafts for the frozen 1.7B target in the same
real cached loop.
Timing includes its audio encoder and prefill. The models share a tokenizer,
and use exact-match greedy verification. The output is target-greedy exact
under exact arithmetic; \cref{app:setup} reports the low-precision check.

\begin{center}
\begin{minipage}{\columnwidth}
\centering
\begingroup\small
\setlength{\tabcolsep}{2pt}
\begin{tabular}{lcc}
\toprule
Set & Full recognizer & Shallow self-draft \\
\midrule
Clean & \MatchedFullCleanSpeed{} & \MatchedSelfCleanSpeed{} \\
Other & \MatchedFullOtherSpeed{} & \MatchedSelfOtherSpeed{} \\
TED-LIUM & \MatchedFullTedSpeed{} & \MatchedSelfTedSpeed{} \\
GigaSpeech & \MatchedFullGigaSpeed{} & \MatchedSelfGigaSpeed{} \\
FLEURS & \MatchedFullFleursSpeed{} & \MatchedSelfFleursSpeed{} \\
\bottomrule
\end{tabular}
\endgroup
\captionof{table}{Speedup over autoregressive decoding on matched slices with
$K{=}4$ and $n{=}120$ per evaluation set. Full-recognizer brackets give paired
$95\%$ bootstrap intervals.}
\label{tab:paired-full-recognizer}
\end{minipage}
\end{center}

\paragraph{Proposal quality and cost.}
The full recognizer reaches restart acceptance \MatchedFullRestart{} and
continuation acceptance \MatchedFullContinuation{}, but adds
\MatchedFullParameters{} parameters and \MatchedFullStaticWeight{} of static
weights. The shallow self-draft uses \MatchedShallowParameters{} trainable
parameters. The full recognizer remains slower than autoregressive decoding on
all five sets, while the shallow self-draft remains faster. Better proposals
alone do not offset the cost of a complete second recognizer in this setup.

\paragraph{Matched timing.}
Both systems are timed serially on the same idle A100-SXM4-40GB. Because the
target dominates peak allocation, we compare the additional static model
weights rather than peak memory.

\subsection{Whisper Cross-Architecture Check}
\label{app:whisper}
\label{app:whisper-feas}

\paragraph{Audio access and continuation.}
We test whether the restart and continuation ordering transfers to Whisper.
A real cached loop uses Whisper-large-v3 as the frozen target and compares
matched one-layer drafts with and without per-step audio cross-attention
~\citep{radford2023whisper,li2025eagle3}. SPGISpeech and VoxPopuli extend the
evaluation beyond the five core sets
~\citep{oneill2021spgispeech,wang2021voxpopuli}.

\begin{table*}[t]
\centering
\begingroup\small
\begin{tabular}{lccc}
\toprule
Set & With audio speedup $\uparrow$ & Without audio speedup $\uparrow$ &
WER $\downarrow$ \\
\midrule
LibriSpeech Clean & $2.75\times$ & $1.03\times$ & $2.5\%$ \\
LibriSpeech Other & $2.31\times$ & $1.01\times$ & $4.7\%$ \\
TED-LIUM & $2.25\times$ & $1.03\times$ & $5.8\%$ \\
SPGISpeech & $2.35\times$ & $0.99\times$ & $4.0\%$ \\
VoxPopuli & $2.04\times$ & $0.97\times$ & $13.5\%$ \\
GigaSpeech & $2.01\times$ & $0.98\times$ & $11.4\%$ \\
FLEURS & $1.59\times$ & $0.93\times$ & $6.3\%$ \\
\bottomrule
\end{tabular}
\endgroup
\caption{Whisper-large-v3 real cached loop with $K{=}12$ and $n{=}100$
per evaluation set. The draft is trained on LibriSpeech only.}
\label{tab:whisper7}
\end{table*}

The draft with audio access is faster on all seven sets. Its matched
acceptance traces also show a modest restart change and a larger continuation
change. The ordering therefore extends to this Whisper implementation.

\paragraph{Offline and real-loop measurement.}
\label{app:whisper-anatomy}
\label{app:whisper-scoreboard}
\label{app:whisper-metric}
\label{app:whisper-capacity}

To test whether fixed-prefix acceptance predicts deployment behavior, we
evaluate the same Whisper draft with offline
scoring, chained survival, and cached decoding. Offline accepted length is
$9.4$, chained survival gives $4.83$, and the real cached loop gives $4.35$.
Restart acceptance is $0.92$ offline and $0.635$ in the loop. Fixed-prefix
scoring therefore overstates the work accepted during deployment.

\paragraph{Remaining restart error.}
The remaining Whisper restart error may come from position or token
information. We compare unrestricted and correct-position attention, then
probe two inputs to the draft against the target token. Restart acceptance is
$0.597$ with unrestricted attention and $0.529$ with the correct window. The
target feature after combining text and audio predicts the token at $0.948$,
while the compact draft state reaches $0.534$. These measurements point to
token information, rather than audio position, as the main limitation in this
restart comparison.

\subsection{Repository Measurements and Published Systems}
\label{app:landscape}

\paragraph{Repository measurements.}
The repository comparison uses a greedy Whisper implementation, except for the
same-family Qwen row. The rows differ in draft length and evaluation sample.
They compare resource and output assumptions rather than methods under
identical conditions.

\begin{table*}[t]
\centering
\begingroup\small
\begin{tabular}{lcccc}
\toprule
Design & Single model & WER-neutral & Uses audio & Speedup $\uparrow$ \\
\midrule
Paired distilled draft & \xmark & \cmark & \cmark & $3.98$--$4.10\times$ \\
Same-family 0.6B recognizer & \xmark & \cmark & \cmark &
\SystemFullRecognizerSpeed{} \\
CTC or encoder draft & \cmark & \cmark & \cmark & $1.53\times$ \\
Text-tuned draft & \cmark & \cmark & \xmark & Approximately $1.5\times$ \\
Whisper-Medusa & \cmark & \xmark & \xmark & $0.85\times$ \\
Assisted generation & \xmark & \cmark & \cmark & $0.87\times$ \\
Draft with audio access & \cmark & \cmark & \cmark & $2.75\times$ \\
\bottomrule
\end{tabular}
\endgroup
\caption{Repository ASR drafting measurements under the Whisper implementation. The
same-family recognizer row uses Qwen.}
\label{tab:positioning}
\end{table*}

\paragraph{Additional model resources.}
The Whisper draft with audio access adds $27$M parameters and reuses the
target encoder output. The paired distilled draft adds $756$M parameters and
runs its own audio encoder and prefill. In bfloat16, their additional weights
occupy approximately $0.05$ and $1.41$ GiB before activations and caches. The
paired draft therefore adds $28\times$ as many draft parameters.

\paragraph{Published systems under original conditions.}
The published results below retain each system's reported resources, output
constraint, hardware, and evaluation protocol.

\begin{table*}[t]
\centering
\begingroup\small
\begin{tabular}{@{}p{2.9cm}p{3.0cm}p{4.5cm}p{4.7cm}@{}}
\toprule
System & Extra resource & Reported output condition & Reported result \\
\midrule
WhisperKit~\citep{orhon2025whisperkit,cheng2024redrafter} &
Recurrent drafter & Reported lossless & $2.24\times$ on v3 and
$1.25\times$ on turbo \\
SpecASR~\citep{specasr2025} & Companion ASR & WER-preserving &
$3.04$--$3.79\times$ \\
IBM CTC speculation~\citep{saon2026selfspec} & CTC branch &
Relaxed with $12\%$ relative WER increase &
$4.4\times$ inverse real-time factor \\
Whisper-Medusa~\citep{segalfeldman2025whispermedusa} & Multiple heads &
Minimal WER impact reported & Approximately $50\%$ latency reduction \\
Token-Map~\citep{tokenmap2025} & N-gram map &
No accuracy loss reported & $1.27$--$1.37\times$ \\
\bottomrule
\end{tabular}
\endgroup
\caption{Published ASR acceleration systems under their original resource,
output, and hardware conditions.}
\label{tab:landscape}
\end{table*}

\FloatBarrier

\end{document}